%% file: draft4.tex
\documentclass[aps,pre,preprint,onecolumn,citeautoscript,superscriptaddress,floatfix]{revtex4-1}  
\input{headersDraft}

\begin{document}

\title{Fractionalized Fermi liquid\\ on the surface of a topological Kondo insulator}
 \author{Alex Thomson}
 \affiliation{Department of Physics, Harvard University, Cambridge, Massachusetts, 02138, USA}
 \author{Subir Sachdev}
 \affiliation{Department of Physics, Harvard University, Cambridge, Massachusetts, 02138, USA}
 \affiliation{Perimeter Institute for Theoretical Physics, Waterloo, Ontario N2L 2Y5, Canada}
 \date{\today}
 \vspace{0.6in}
\begin{abstract} 
We argue that topological Kondo insulators can also have `intrinsic' topological order associated with fractionalized excitations on their surfaces.
The hydridization between the local moments and conduction electrons can weaken near the surface, and this enables the local moments to
form spin liquids. This co-exists with the conduction electron surface states, realizing a surface fractionalized Fermi liquid. We present mean-field solutions 
of a Kondo-Heisenberg model which display such surfaces. 
\end{abstract}

\maketitle

\section{Introduction}

An important development of the past decade has been the prediction and discovery of topological insulators (TI) \cite{hasan10,qi11,qsh_kane05,bhz,3dTI_Moore07,3dTI_Fu07,3dTI_Roy09}. 
These materials are well-described by traditional band theory, but possess strong spin-orbit interactions that result in a non-trivial winding of the ground state wavefunction in a manner analogous to the integer quantum Hall effect. Since their discovery, the multitudinous effects of interactions have been a prominent topic of study. One compelling proposal to emerge is the notion of a topological \emph{Kondo} insulator (TKI) \cite{dzero10,dzero12,coleman_arcmp}.
In contrast to a band insulator, a Kondo insulator only develops an insulating gap at low temperatures, and the magnitude of the gap is controlled
by electron-electron interactions. 
Doniach explained this phenomenon through the Kondo lattice model \cite{doniach77} in which a lattice of localized moments is immersed within a sea of conduction electrons. At high temperatures, RKKY-type exchange interactions dominate and an ordered magnetic state results. Conversely, at low energies, strong interactions between localized moments and conduction electrons become important; the system crosses over into either a metallic phase well-described by Fermi liquid theory (FL) or, if the chemical potential is appropriately tuned, a Kondo insulator. 
As strong spin-orbit coupling is often present in these materials, the possibility that a Kondo insulator may have a nontrivial topological character is well-justified.

Of specific interest has been the Kondo insulator samarium hexaboride (SmB$_6$).
A number of experiments have examined the proposal that it is a TKI: transport measurements have established the presence of metallic surface states \cite{trans_Wolgast13,trans_Kim12,trans_Zhang13,trans_Phelan14,trans_Kim13}, and angle-resolved photoemission spectroscopy (ARPES) results appear consistent with the expected Dirac surface cones \cite{arpes_Neupane13,arpes_Jiang15,arpes_Xu13,arpes_Denlinger14,arpes_XuSpin}. 
Nonetheless, the spin-polarized ARPES measurements \cite{arpes_XuSpin} remain controversial.

However, as the TKI phase is well-described within a mean field framework \cite{coleman_arcmp}, its topological properties are not expected to be markedly different from what has already been observed in its uncorrelated cousins.
More intriguing is the potential the topologically protected surface states present for new interesting phases  \cite{roy14,nikolic14,alexandrov15,iaconis15}. 
In SmB$_6$, this expectation is motivated experimentally by ARPES measurements which find light surface quasiparticles \cite{arpes_Neupane13,arpes_Xu13,arpes_Jiang15} in contradiction to current theories which predict heavy particles at the surface \cite{dzero10,dzero12,alexandrov13,lu13}.  Ref.~\cite{alexandrov15} proposes  ``Kondo breakdown" at the surface as an explanation. They show that the reduced coordination number of the localized moments at the surface may lead to a suppressed Kondo temperature. At low temperature these moments are thermally decoupled from the bulk.

In this paper, we propose the existence of a fractionalized Fermi liquid (SFL$^*$) on the surface of a TKI.
This state is characterized by ``{\em intrinsic\/} topological order'' on the surface of a TKI, in which the local moments form a spin liquid
state which has `fractionalized' excitations with quantum numbers which cannot be obtained by combining those of one or more 
electrons \cite{balents_review}. Rather than being thermally liberated, as in Ref.~\onlinecite{alexandrov15}, the surface local moments exploit their mutual
exchange interactions to decouple from the conduction electrons, and form a spin liquid state, as in the fractionalized Fermi liquid state (FL$^*$) \cite{senthil03,senthil04}. We will present mean-field computations on a Kondo-Heisenberg lattice model which demonstrate the formation of mutual singlets
between the surface local moments, while conducting surface states of light electronic quasiparticles are also present.

Somewhat confusingly, our SFL$^*$ state is `topological' in two senses of the word, a consequence of unfortunate choices (from our perspective) 
in the conventional terminology.
As in conventional TI, it is `topological' because it has gapless electronic states on the surface induced by the nature of the bulk band structure.
However, it is also `topological' in the sense of spin liquids \cite{balents_review}, because of the presence of fractionalized excitations among
the local moments on the surface. 

The outline of our paper is as follows. We specify our Kondo-Heisenberg model in Section~\ref{sec:model}. In Section~\ref{sec:trans}, we present the mean-field solution
of this model for the case of a translationally-invariant square lattice with periodic boundary conditions. The effect of the surface on the mean field solutions is addressed in Section~\ref{sec:boundary} where the presence of the SFL$^*$ state is numerically demonstrated. 
We conclude in Section~\ref{sec:discussion} with a discussion of our results and their relevance to physical systems.

\section{Model}
\label{sec:model}

Here we present the specific form of the Kondo-Heisenberg lattice model to be studied:
\eq{H=H_c+H_H+H_K \label{eqn:Htot}\,.}
The first terms represents the hopping Hamiltonian of the conduction electrons,
\eq{\label{eqn:condterm}
H_c&= -\sum_{\braket{ij}} t_{ij}\( c_{i\a}^\dag c_{j\a} + h.c.\) \,,
}
where the operator $c_{i\a}^\dag$ creates an electron at site $\vr_i$ of spin $\a=\ua,\da$. 
The remaining two terms establish the form of the interactions: $H_H$ is a generalized Heisenberg term which specifies the inter-spin interaction while $H_K$ is a Kondo term and describes the electron-spin exchange. 


The spin-orbit coupling of the $f$-orbital imposes a classification in terms of a $(2J+1)$ multiplet, where $J$ is the total angular momentum. In general, this degeneracy is further lifted by crystal fields and we will consider the simplest case of a Kramers degenerate pair of states. 
We start from an Anderson lattice model \cite{anderson61} with hopping $t_f$ between $f$-orbitals and onsite repulsion $U_f$. To access the Kondo limit ($U_f\rightarrow\infty$), we perform a Schrieffer-Wolff transformation \cite{schriefferwolff} and obtain a term of the form
\eq{
H_H = -{J_H\o2}\sum_{\braket{ij}}f_{i\a}^\dag f_{j\a} f_{j\b}^\dag f_{i\b}\,,
}
where $f_{i\a}$ creates a spinon at site $\vr_i$, and $J_H\sim t^2_f/U_f$. This limit imposes the constraint  $\sum_\a f_{i\a}^\dag f_{i\a} = 1 $ and further ensures that the correct commutation relations for the ``spin" operators $S^a_j = {1\o 2} f^\dag_{j\a}\s^a_{\a\b}f_{j\b}$ are obeyed. By using the Fierz identity (and dropping a constant) we can verify that we indeed have the familiar Heisenberg term:
\eq{
H_H={J_H\o4}\sum_{\braket{ij}} f_{i\a}^\dag \bb{\s}_{\a\b} f_{i\b}\cdot f_{j\g}^\dag \bb{\s}_{\g\d} f_{j\d}=J_H\sum_{\braket{ij}}\v{S}_i\cdot\v{S}_j\,,
\label{eqn:heisterm}
}
where $\bb{\s}=\(\s^x,\s^y,\s^z\)$.
It is important to note that the spinon operators $f_{j\a}$ do not have a uniquely defined phase. In fact, by choosing to represent the spins in terms of constrained fermion operators, we are formulating the Kondo lattice model as a U(1) gauge theory. This emergent gauge structure is what permits a realization of the fractionalized phases we will discuss \cite{senthil03,senthil04}.

For the electron-spin interaction, $H_K$, we follow the construction of Coqblin-Schrieffer \cite{coqblinschrieffer} for systems with spin-orbit coupling. In order for the interaction to transform as a singlet, the electron and spin must couple in a higher angular momentum channel.
For simplicity, we assume a square lattice and that the spins and conduction electrons carry total angular momentum differing by $l=1$. In the Anderson model, an appropriate interaction term is then
\eq{\label{eqn:spinElInt}
H_{int}\sim \V \sum_{\vk,\a} \(\a\sin k_x - i\sin k_y\) c_{\vk\a}^\dag\ket{f^0}\bra{f^1;\a} +h.c.
}
For instance, the interaction between moments with total angular momentum $J=3/2$ and spin-1/2 electrons would take this form.
We will verify in the next section that for the purpose of obtaining a TKI, this coupling is sufficient.
We next define the electron operators
\eq{\label{eqn:pwave}
d_{\vk\a}&=2\( \a\sin k_x +i\sin k_y \) c_{\vk\a}, & d_{i\a}&=-i\a(c_{i+\v{\hat{x}},\a}-c_{i-\v{\hat{x}},\a})+(c_{i+\v{\hat{y}},\a}-c_{i-\v{\hat{y}},\a})\,.
}
and, taking the same $U_f\rightarrow\infty$ limit as above, again implement the Schrieffer-Wolff transformation \cite{schriefferwolff} to obtain
\eq{
H_K = -{J_K\o4} \sum_i f^\dag_{i\a}d_{i\a}d_{i\b}^\dag f_{i\b} 
}
where $J_K\sim \V^2/U_f$.

We next perform a Hubbard-Stratonovich transformation of the Kondo and Heisenberg terms:
\eq{
H'=& H_1 + H_0\nt
 H_1=&-\sum_{\braket{ij}}\( (t_{ij}-\d_{ij}\m_i) c_{i\a}^\dag c_{j\a} + h.c.\)  + {1\o2}\sum_{j\a} \[V_{j} f_{j\a}^\dag d_{ j\a} + V_{j}^* d_{ j\a}^\dag f_{j\a}\] 
\nt
&-{1\o2}\sum_{j\a\hvm}\[\chi_{j\m} f_{j+\hvm,\a}^\dag f_{j\a} + \chi^*_{j\m} f_{j\a}^\dag f_{j+\hvm,\a} \] 
+ \sum_j \lam_j f_{j\a}^\dag f_{j\a} 
\nt
 H_0=&\sum_j\[-\lam_j+{\abs{V_{j}}^2\o J_K}+\sum_{\hvm} {\abs{\chi_{j\m}}^2\o 2J_H}\]\,.
\label{eqn:MFham}
} 
We proceed with a saddle-point approximation, and treat the fields $V_j$, $\chi_{j\m}$, and $\lam_j$ as real constants subject to the self-consistency conditions
\eq{
V_j&=-{J_K\o2}\Braket{d_{j\a}^\dag f_{j\a}}, & \chi_{j\m}&={J_H} \Braket{f_{j\a}^\dag f_{j+\hvm,\a}}
\label{eqn:MFVChi}\,,\\
1&=\Braket{f_{j\a}^\dag f_{j\a}}\,.
\label{eqn:MFLam}
}
This can be formally justified within a large-$N$ expansion of Eq.~\ref{eqn:Htot}, with $N$ the number of spinons.
As we are specifying to the case of an insulator, it further makes sense to require perfect half-filling. Since $n_f=1$ already, this results in a final equation for the chemical potential $\m_j$:
\eq{
1&=\Braket{c_{j\a}^\dag c_{j\a}} \,.
\label{eqn:MFMu}
}

\section{Translationally invariant system}
\label{sec:trans}

We begin by solving Eqs.~\ref{eqn:MFVChi}$\,-\,$\ref{eqn:MFMu} in a translationally invariant system with periodic boundary conditions on a square lattice. Letting $V_j=V$, $\chi_{jx}=\chi_{jy}=\chi$, $\lam_j=\lam$ and $\mu_j=\mu$, we perform a Fourier transform:
\eq{\label{eqn:hamtrinv1}
H_1&=\sum_\vk \Psi_\vk^\dag \mathcal{H}(\vk) \Psi_\vk
\qquad\qquad & \Psi^\dag_\vk&=\( c_{\vk\ua}^\dag, f_{\vk \ua}^\dag ,c_{\vk\da}^\dag,f_{\vk \da}^\dag\)\\
\mathcal{H}(\vk)
&=\begin{pmatrix}
h(\vk) & 0 \\
0 & h^*(-\vk) 
\end{pmatrix} & 
h(\vk)&=\begin{pmatrix}
\ep_c(\vk) & V(\sin k_x +i\sin k_y)\\
V(\sin k_x -i\sin k_y) & \ep_f(\vk) \\
\end{pmatrix}\,. \label{eqn:hamtrinv2}
}
For simplicity, we only consider nearest-neighbour coupling between spins; for the electron dispersion, a slightly more general description is required and we also take next-nearest neighbour hopping into account. The dispersions are given by
\eq{
\ep_c(\vk) &= - t_1(\cos k_x +\cos k_y ) - 2 t_2 \cos k_x \cos k_y -\mu,& \ep_f(\vk) & =-\chi(\cos k_x + \cos k_y) +\lam
\label{eqn:dispersions}
}
where the subscripts ``$c$" and ``$f$" refer to the electrons and spinons respectively. In the following, we will use units of energy where $t_1=1.0$.

Since TI's exist as a result of a band inversion, it's important to ask which sign $\chi$ will take.
Naturally, when $V=0$, the particle-hole symmetry of our mean field \emph{ansatz} implies that $\chi>0$ and $\chi<0$ have the same energy. At finite hybridization, however, one will become preferable.
We note that when $\chi$ and $t_1$ have opposite signs, the energy of the lower band will be less than the Fermi energy and hence occupied throughout most of the Brillouin zone (BZ): an increase in $V$ will push most of these states to lower energies.  Conversely, if $\chi$ and $t_1$ take the same sign, in one of part of the BZ  no states will lie below the Fermi energy while in another both the upper and lower band will. It therefore makes sense to expect $\text{sign}\(\chi\)=-\text{sign}\(t_1\)$. In the parameter regime explored, the numerics always find this to be the case.

By construction, the Hamiltonian $H_1$ supports a non-trivial topological phase and is in fact the familiar Bernevig-Zhang-Hughes model \cite{bhz} used to describe the quantum spin Hall effect in HgTe wells.
We can see this by studying the eigenfunctions of $h(\vk)$: 
\eq{
\psi_\pm(\vk)={1\o\sqrt{2d(d\pm d_3)}}\begin{pmatrix} d_3 \pm d \\ V(\sin k_x + i \sin k_y) \end{pmatrix}
}
where $d(\vk)=\sqrt{d_3(\vk)^2+V^2(\sin^2 k_x + \sin^2 k_y) }$, and $d_3(\vk)=\(\ep_c(\vk)-\ep_f(\vk)\)/2$. If {$d_3(\vk)>0$} or {$d_3(\vk)<0$} for all $\vk$, these functions are well-defined on the entire BZ and the system is in a topologically trivial phase \cite{bernevig}. If this is not the case then it is impossible to choose a globally defined phase -- the ground state wavefunction has nontrivial winding and characterizes a topological insulator. From Eq.~\ref{eqn:dispersions}, we see that this occurs when 
\eq{
-2 < { \m+\lambda+2t_2\o t_1-\chi } <2\,.
\label{eqn:TIcond}}
Alternatively, we can obtain the same result by calculating the $\Zt$ invariant $\n$ \cite{fukane}: when Eq.~\ref{eqn:TIcond} holds, $\n=-1$ and the system is a TI. 

We will typically be studying systems with $\abs{t_1}\gg \abs{\chi}$ and $\abs{t_2/t_1}$ small (implying $\mu$ and $\lam$ small as well), so Eq.~\ref{eqn:TIcond} is not difficult to fulfill. In Fig.~\ref{fig:specVChiConst} the energy spectrum of the system in a slab geometry is shown for $J_H=0.15$, $J_K=0.3$. Half-filling is maintained on every site (see Appendix~\ref{app:MFbound}), but $V$ and $\chi$ were determined by self-consistently solving Eq.~\ref{eqn:MFVChi} in a periodic system. In Fig.~\ref{fig:specVChiConst}(b), the topologically protected Dirac cone is clearly visible.

\newcommand{\incgrC}[1]{\includegraphics[scale=0.47]{#1}}
\begin{figure}
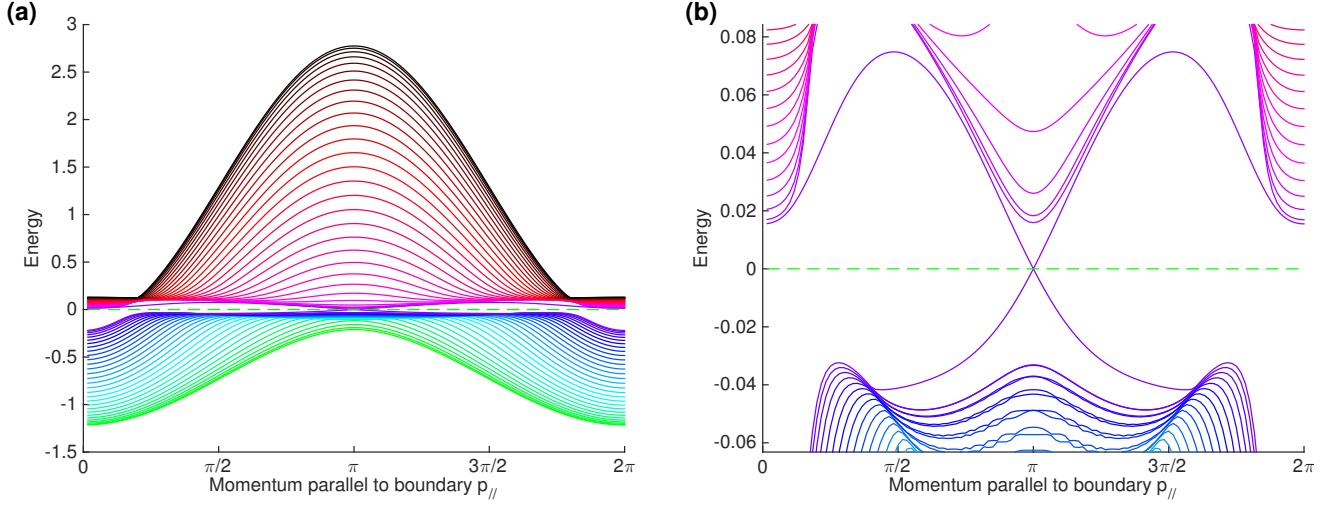

\centering
\incgrC{{Figures/constVChiFullSpec}.pdf}
\hspace{4mm}
\incgrC{{Figures/constVChiDirac}.pdf}
\caption{Energy spectrum for $J_H=0.15$, $J_K=0.3$. Both $V$ and $\chi$ are constant throughout the bulk, but both $\mu_i$ and $\lam_i$ have been self-consistently solved to ensure that $n_c=n_f=1$ on every site (see Appendix~\ref{app:MFbound}). (a) The full spectrum is shown. (b) A closer view of the insulating gap, where the Dirac cone is clearly visible. We use units with $t_1=1.0$. Calculations were done with $t_2=-0.25$ and at a temperature of $10^{-5}$. }
\label{fig:specVChiConst}
\end{figure}

If we ignore the effect the boundary will have on the values of $V$, $\chi$, and $\lam$, we can calculate the Fermi velocity of the Dirac cone \cite{bernevig}:
\eq{\label{eqn:vFCalc}
v_F = 2V \sqrt{\frac{\abs{ \chi  (t_1-2t_2)}}{\abs{t_1-2t_2}+\abs{\chi}} }\sim 2V\sqrt{\abs{\chi}} 
}
where we've assumed $\abs{\chi}\ll t_1$ in the second equation. This is consistent with the prediction that the quasiparticles at the surface be heavy \cite{dzero10,dzero12,alexandrov13,lu13}. For the parameters shown in Fig.~\ref{fig:specVChiConst}, this formula predicts $v_F= 0.0592$, consistent with the numerically determined value $v_F=0.0585$.

\section{System with boundary}
\label{sec:boundary}

We now consider the effect the boundary will have on the mean field configuration and demonstrate the presence of two new fractionalized phases. Generally,
 we expect that the lower coordination number at the boundary will suppress the (nonlocal) hybridization: $V_{\rm surf}\sim 3V_{\rm bulk}/4$. While the decrease in $V_{\rm surf}$ will induce an increase in the spinon bond parameter $\chi_{i\m}$ \cite{paul08} both parallel  and perpendicular to the surface, the parameter parallel to the surface will be more strongly affected. Since Heisenberg coupling ultimately favours an alternating bond order, in the absence of hybridization $V$, this anisotropy will result in a further decrease in the magnitude of the spinon bond parameter perpendicular to the surface, $\abs{\chi_\perp}$. 
 
\begin{figure}
\centering
\includegraphics[scale=0.65]{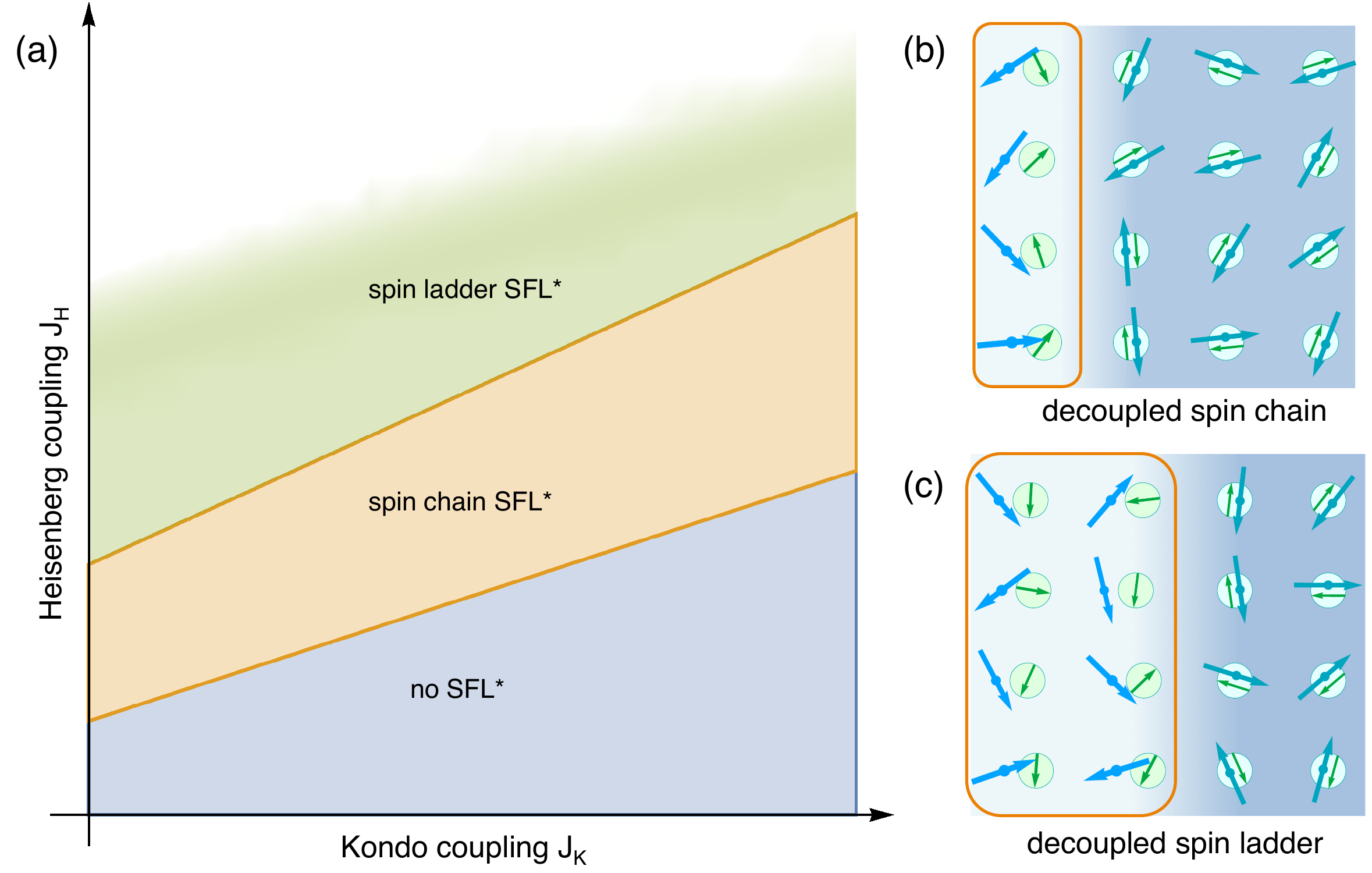}
\caption{(a) Schematic phase diagram of surface states. (b),(c) Cartoon depictions of surface FL$^*$ states. In the dark blue region, the electron spins and localized moments are locked into singlets. Towards the edge (the pale blue region outlined in orange) the conduction
electrons decouple from the moments, and the latter form a spin liquid. Naturally, the conduction electrons remain coupled to each other at all sites.}
\label{fig:phDiag}
\end{figure}

When these effects are predominant, an FL$^*$ on the surface is realized: the hybridization $V_i$ vanishes on one or more layers at the surface and $\chi_\perp$ vanishes on the innermost layer. 
The existence of the SFL$^*$ phase is shown numerically by self-consistently solving Eqs.~\ref{eqn:MFVChi}$-$\ref{eqn:MFMu}  in a slab geometry and comparing ground states energies (some details are given in Appendix~\ref{app:MFbound}). The resulting phase diagram is shown in Fig.~\ref{fig:phDiag}(a). In fact, we find two distinct SFL$^*$ phases: a decoupled spin chain and a decoupled spin ladder, which are depicted in Figs.~\ref{fig:phDiag}(b) and \ref{fig:phDiag}(c) respectively.
In Fig.~\ref{fig:MFparams} we plot the spatial dependence of the mean field parameters in both SFL$^*$ states. The plots in the left column correspond to a spin chain SFL$^*$ state whereas the right column corresponds to a spin ladder SLF$^*$ state. The phases are distinguished by whether $V$ vanishes on the first site only or on both the first and second site, shown in Fig.~\ref{fig:MFparams}(a) and (b) respectively. In Figs.~\ref{fig:MFparams}(c) and (d) our intuition regarding the behaviour of $\chi$ near the boundary is confirmed: $\abs{\chi_\perp}$ is suppressed to zero whereas $\abs{\chi_{\parallel}}$ increases to the value it would assume in a single dimension. The fluctuations of the Lagrange multiplier field $\lambda$ (Figs.~\ref{fig:MFparams}(e) and (f)) are a reflection of the on-site requirement of half-filling for both the spinons and electons.

\newcommand{\incgrB}[1]{\includegraphics[scale=0.65]{#1}}

\newcommand{\incgrD}[1]{\includegraphics[scale=0.47]{#1}}
\begin{figure}
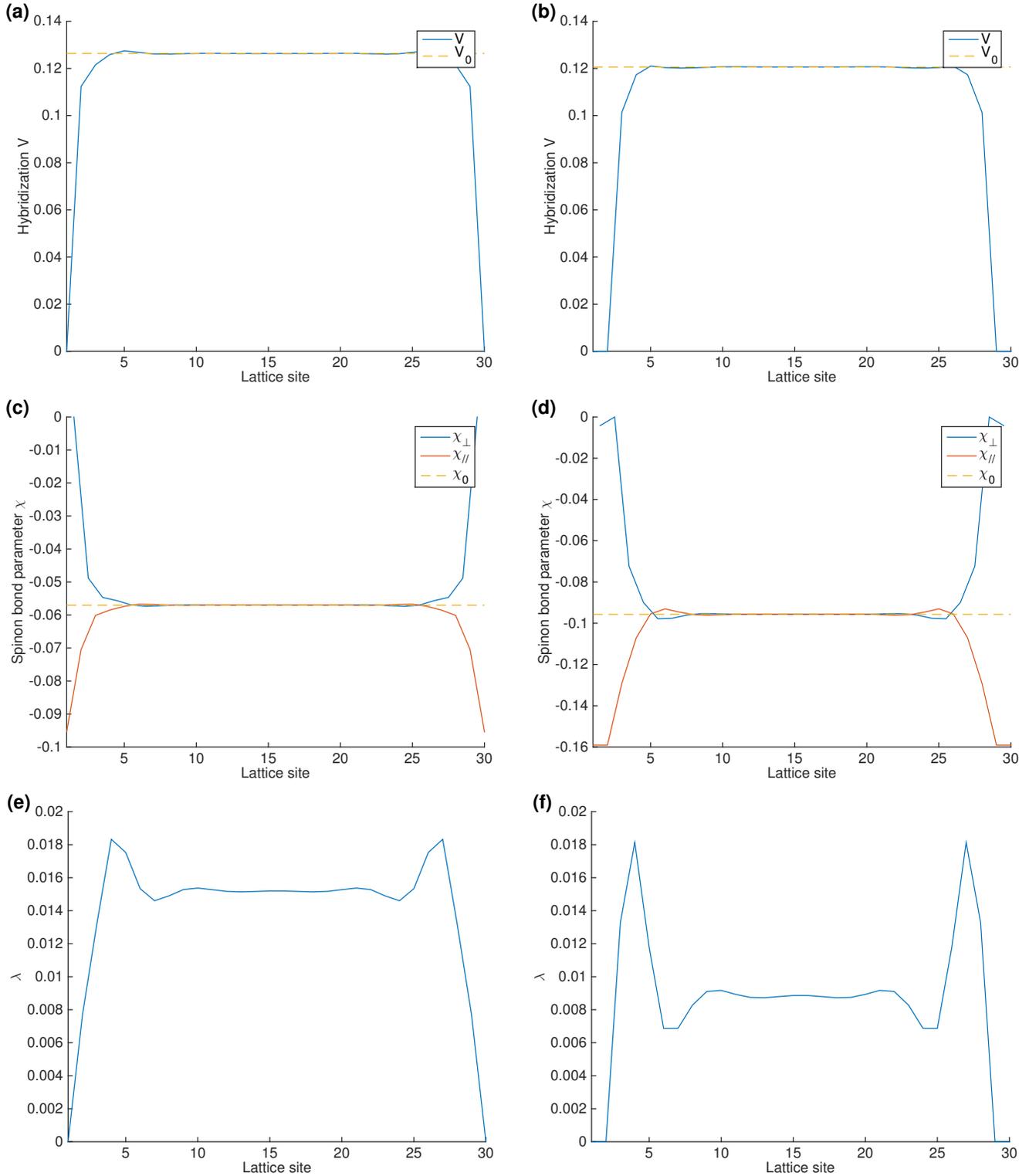

\centering
\incgrD{{Figures/VX/hyb_lab}.pdf}
\hspace{4mm}
\incgrD{{Figures/V2X2/hyb_lab}.pdf}
\\\vspace{2mm}
\incgrD{{Figures/VX/chi_lab}.pdf}
\hspace{4mm}
\incgrD{{Figures/V2X2/chi_lab}.pdf}
\\\vspace{2mm}
\incgrD{{Figures/VX/lambda_lab}.pdf}
\hspace{4mm}
\incgrD{{Figures/V2X2/lambda_lab}.pdf}
\caption{Spatial dependence of mean field parameters in SFL$^*$ phases. In the left column, we plot values corresponding to the spin chain SFL$^*$ ($J_H=0.15$, $J_K=0.3$) while on the right values corresponding to the spin ladder SFL$^*$ ($J_H=0.25$, $J_K=0.3$) are shown. (a),(b)~Hybridization $V_i$. (c),(d)~Spinon bond parameters $\chi_{i\m}$ in the direction perpendicular (blue) and parallel (red) to the boundary. (e),(f)~The Lagrange multiplier field $\lambda_i$. In (a)$-$(f), the yellow dashed line plots the value obtained in the translationally invariant case. We use units with $t_1=1.0$. Calculations were done with $t_2=-0.25$ and at a temperature of $10^{-5}$.}
\label{fig:MFparams}
\end{figure}

\newcommand{\incgrA}[1]{\includegraphics[scale=0.47]{#1}}
\begin{figure}
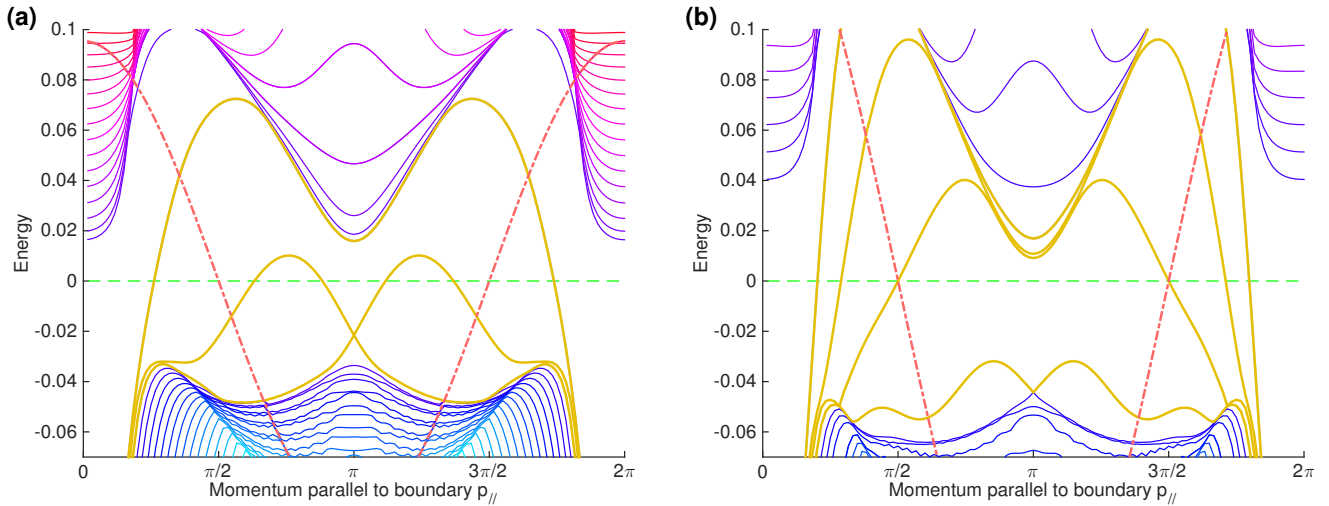

\centering
\incgrA{{Figures/spinChainSpec}.pdf}
\hspace{4mm}
\incgrA{{Figures/spinLadderSpec}.pdf}
\caption{Energy spectra in SFL$^*$ phases. (a) Spin chain SFL$^*$ ($J_H=0.15$, $J_K=0.3$). The ground state has $V_i=0$ on the first surface layer and the moments form a spin chain decoupled from the bulk.  (b) Spin ladder SFL$^*$ ($J_H=0.25$, $J_K=0.3$). The ground state has $V_i=0$ on the first two layers and a spin ladder is present on the surface. In both figures, the dash-dotted red curve represents the one-dimensional cosine dispersion found for the spinons and is merely an artifact of the \emph{ansatz}. We use units with $t_1=1.0$. Calculations were done with $t_2=-0.25$ and at a temperature of $10^{-5}$.}
\label{fig:specChainLadder}
\end{figure}

In Fig.~\ref{fig:specChainLadder}(a), the spectrum of the spin chain SFL$^*$ state is shown. The red dash-dotted curve is the dispersion of the spinons calculated at mean field. While we do not claim that this accurately represents the Heisenberg chain, we nonetheless expect gapless spin excitations \cite{haldane}. The remaining in-gap states can be understood as the result of the mixing of the surface layer of conduction electron with the Dirac cone. Consistent with its topology, even if the Dirac cone is no longer present at the chemical potential, two chiral bands traverse the gap from the conduction to the valence band and the surface is metallic. In this case, an additional four metallic surface states per spin are present, but these are not topologically protected and we can imagine pushing them below the chemical potential in a number of ways, such as, for instance, softening the restriction imposed by Eq.~\ref{eqn:MFMu}. 

The spectrum corresponding to the second surface FL$^*$ state, the decoupled spin ladder, is shown in Fig.~\ref{fig:specChainLadder}(b). The red curve representing the spinons is now two-fold degenerate per spin (a small splitting is hidden by the thickness of the line). Even more so than for the spin chain, this result is an artifact of the mean field calculations: save in the limit where the legs of the ladder are completely decoupled, a ladder of spin-1/2 particles is gapped \cite{haldane,giamarchi03}. 

In both phases, the metallic bands have lighter quasiparticles than predicted by the translationally invariant theory in Eq.~\ref{eqn:vFCalc}. For the spin chain, the surface velocity of the leftmost state in Fig.~\ref{fig:specChainLadder}(a) is $v_F=0.095$, compared to $v_F=0.052$ for the Dirac cone of Fig.~\ref{fig:specVChiConst}(b).
For the spin ladder, the effect is even more pronounced. There, the lightest state has a Fermi velocity of $v_F= 0.48$ compared to the translationally-invariant value $v_F= 0.072$.

The structure of the fractionalized excitations in the SFL$^*$ states found here is rather simple: just a free gas of neutral $S=1/2$ spinon excitations.
We view this mainly as a `proof of principle' that such SFL$^*$ states can exist on the surface of TKI. Clearly, more complex types of spin liquid states
are possible on the surface, and also in three-dimensional TKI with two-dimensional surfaces.

\section{Discussion}\label{sec:discussion}

The strong electron-electron interactions in topological Kondo insulators make them appealing candidates for searching for novel correlated
electron states. In many heavy-fermion compounds, the strong interactions acting on the $f$-electron local moments are quenched
by the Kondo screening of the conductions electrons, and the resulting state is eventually a Fermi liquid, or a band insulator for suitable density.
The topological Kondo insulators offer the attractive possibility that the hybridization between the local moments and the conduction
electron states can be weakened near the surface \cite{roy14,alexandrov15}, and this could explain the light effective masses associated
with the surface electronic states \cite{arpes_Neupane13,arpes_Xu13,arpes_Jiang15}. With the weakened hybridization, we have proposed here that the
local moments may form a spin liquid state with `intrinsic' topological order. As the fractionalized excitations of such a spin liquid co-exist
with the conduction electron surface states similar to those of a conventional TI, the surface realizes a fractionalized Fermi liquid \cite{senthil03,senthil04}.

This paper has presented mean-field solutions of Kondo-Heisenberg model on a square lattice 
which act as a proof-of-principle of the enhanced stability of the such
surface fractionalized Fermi liquids (SFL$^*$). We fully expect that such solutions also exist on the surfaces of 
three-dimensional lattices, relevant to a Kondo insulator like SmB$_6$. 

The recent evidence for \emph{bulk} quantum oscillations in insulating SmB$_6$ \cite{QO_suchitra15}  is exciting evidence for the 
non-trivial many-electron nature of these materials. It has been proposed \cite{Cooper15} that these oscillations appear
because the magnetic-field weakens the hybridization between the conduction electrons and the local moments, and this releases
the conduction electrons to form Fermi surfaces leading to the quantum oscillations. The fate of the local moments was not discussed
in Ref.~\onlinecite{Cooper15}, but a natural possibility is that they form a bulk spin liquid, similar to the surface spin liquid we have
discussed here. Thus, while we have proposed here the formation of a SLF* states in SmB$_6$ in zero magnetic field, it may well be that a bulk FL$^*$ 
state forms in high magnetic field.

\section*{Acknowledgments} 
We thank D.~Chowdhury for significant discussions at the early stages of this project.
We also thank S.~Sebastian and J.~D.~Sau for helpful discussion.
AT is supported by NSERC.
This research was supported by the NSF under Grant DMR-1360789.
Research at Perimeter Institute is supported by the Government of Canada through Industry Canada 
and by the Province of Ontario through the Ministry of Research and Innovation.   
 
\appendix

\section{Mean field theory with boundary} 
\label{app:MFbound}

In this appendix, we consider the mean field equations in the presence of a boundary. We define the lattice to be finite in the $x$-direction, $x_i=1,\ldots,N$, and infinite in the $y$-direction (to remove factors of $i$, we actually switch the $x$- and $y$-directions compared to Eq.~\ref{eqn:pwave}). We rewrite the Fourier transform, which is now only valid in the $y$-direction:
\eq{\label{eqn:fourtrans}
c_{ij\s}&=\int {dk\o2\pi} e^{iky_j}c_{ik\s} ,
& 
d_{ij\a}&=\int{dk\o2\pi}e^{iky_j}\[2\a\sin k \, c_{ik\a} + (c_{j+1,k\a}-c_{i-1,k\a}) \] ,\nt
f_{ij\s}&=\int{dk\o2\pi} e^{iky_j}f_{ik\s} .
}

Translational invariance in the $y$-direction implies that the mean field parameters will depend only on the distance from the boundary -- the roman indices $i$, $j$, etc. label the $x$-coordinate only. 
We express the Hamiltonian in block form as
\eq{
\ham_{MF}&=\sum_k \Psi^\dag_{k\s} \hamk_\s(k)\Psi_{k\s}, 
& 
\Psi_{k\s}^\dag &= \( c_{1k\s}^\dag, c_{2k\s}^\dag, \hdots, f_{1k\s}^\dag, f_{2k\s}^\dag,\hdots\)
\nt
& & &=
\(\psi_{1k\s}^\dag,\psi_{2k\s}^\dag,\hdots,\psi_{1+N,k\s}^\dag,\psi_{2+N,k\s}^\dag,\hdots\)
\nt
\hamk_\ua(k) & = \begin{pmatrix}
h_c(k) & h_{cf}(k) \\
h_{cf}^\dag(k) & h_f(k)
\end{pmatrix}
&
\hamk_\da(k)&=\hamk_\ua(-k)^*
}
with blocks given by 
\begin{flalign}
h_c(k) &=
-{t_1\o2}\begin{pmatrix}
  2\cos k & 1 & 0 & \cdots \\
  1 & 2 \cos k & 1 &\cdots \\
  0 & 1 & 2\cos k  &\cdots \\
  \vdots & \vdots & \vdots & \ddots
  \end{pmatrix}
  - t_2 \cos k
\begin{pmatrix}
  0 & 1 & 0 & \cdots \\
  1 & 0 & 1 &\cdots \\
  0 & 1 & 0 &\cdots \\
  \vdots & \vdots & \vdots & \ddots
  \end{pmatrix}
  -
  \begin{pmatrix}
   \m_1 & 0 & 0 & \cdots \\
  0 & \m_2 & 0 &\cdots \\
  0 & 0 & \m_3 &\cdots \\
  \vdots & \vdots & \vdots & \ddots
  \end{pmatrix}&
\end{flalign}
\begin{flalign}
h_f(k)&= 
-{1\o2}\begin{pmatrix}
  2\chi_{1y}\cos k &\chi_{1x} & 0 & 0 &\cdots &\cdots&\cdots \\
  \chi_{1x} & 2\chi_{2y} \cos k & \chi_{2x}  & 0 & \cdots &\cdots&\cdots\\
  0 & \chi_{2x} &2 \chi_{3y}\cos k  & \chi_{3x} &\cdots&\cdots&\cdots \\
  \vdots & \vdots & \vdots & \vdots & \ddots &\cdots&\cdots \\
   \vdots & \vdots & \vdots & \vdots & \vdots &2\chi_{N-1,y}\cos k &\chi_{N-1,x} \\
    \vdots & \vdots & \vdots & \vdots & \vdots &\chi_{N-1,x}&2\chi_{Ny} \cos k 
  \end{pmatrix}
  \nt
  &\qquad+ \begin{pmatrix}
  \lam_1 & 0 & 0 & 0 &\cdots &\cdots&\cdots \\
  0 & \lam_2 & 0 & 0 &\cdots &\cdots&\cdots \\
  0 & 0 & \lam_3 & 0 &\cdots &\cdots&\cdots \\
  \vdots & \vdots & \vdots & \vdots & \ddots &\cdots&\cdots \\
   \vdots & \vdots & \vdots & \vdots & \vdots &\lam_{N-1} &0 \\
    \vdots & \vdots & \vdots & \vdots & \vdots &0&\lam_{N}
    \end{pmatrix}&
\end{flalign}
\begin{flalign}
 &h_{cf}(k)={1\o2}
 \begin{pmatrix}
  2V_{1}\sin k & -V_{2} & 0 & \cdots \\
  V_{1} & 2V_{2}\sin k & -V_{3} &\cdots \\
  0 & V_{2} & 2V_{3}\sin k &\cdots \\
  \vdots & \vdots & \vdots & \ddots
  \end{pmatrix} \,.&
\end{flalign}
To determine correlation functions, we diagonalize the Hamiltonian numerically. For each $k$, we find the matrices $U(k)$ such that
\eq{
U^\dag \hamk(k) U(k) &= \Lambda(k), & \Lambda_{ij}(k) &= \d_{ij} E_j (k) \,.
}
Then, the mean field equations of Eqs.~\ref{eqn:MFVChi}$\,-\,$\ref{eqn:MFMu} may be expressed as
\eql{\label{eqn:nfncMFbound}
&1= \sum_\a\int{dk\o2\pi}\Braket{f_{kj\a}^\dag f_{kj\a}} = \sum_\a\int{dk\o2\pi}\Braket{\psi_{k,j+N,\a}^\dag\psi_{k,j+N,\a}} = 2\int {dk\o2\pi}\sum_{l=1}^{2N}n(E_l(k))U_{j+N,l}(k)U^\dag_{l,j+N}(k) &\\
&1= \sum_\a\int{dk\o2\pi}\Braket{c_{kj\a}^\dag c_{kj\a}} = \sum_\a\int{dk\o2\pi}\Braket{\psi_{kj\a}^\dag\psi_{kj\a}} = 2\int {dk\o2\pi}\sum_{l=1}^{2N}n(E_l(k))U_{jl}(k)U^\dag_{lj}(k)&
}
\eql{
&V_i  = -{J_K\o2}\sum_\a \int{dk\o2\pi}\Braket{d^\dag_{ik\a}f_{ik\a}} 
=
-{J_K\o 2}\sum_\a \int {dk\o2\pi}\[ 2\a \sin k \Braket{c_{ik\a}^\dag f_{ik\a}} + \Braket{c_{i+1,k\a}^\dag f_{ik\a}}-\Braket{c_{i-1,k\a}^\dag f_{ik\a}} \]&
\nt
&\ph{V_i}= -{J_K\o2}\sum_\a\int{dk\o2\pi}\[2\a\sin k \Braket{\psi_{ik\a}^\dag \psi_{i+N,k\a}} +\Braket{\psi^\dag_{i+1,k\a}\psi_{i+N,k\a}}-\Braket{\psi^\dag_{i-1,k\a}\psi_{i+N,k\a}}\] \nt
&\ph{V_i}= -J_K \int {dk\o2\pi} \sum_{l=1}^{2N} n(E_l(k))\[2\sin k \;U_{i+N,l}(k)U^\dag_{l,i}(k)+U_{i+N,l}(k)U^\dag_{l,i+1}(k)-U_{i+N,l}(k)U^\dag_{l,i-1}(k)\] &
}
\eql{\label{eqn:chiMFbound}
&\chi_{ix} = {J_H\o2} \sum_\a  \int{dk\o2\pi} \[ \Braket{f^\dag_{i+1,k\a}f_{ik}}+\Braket{f_{ik\a}^\dag f_{i+1,k\a}} \] 
\nt
&\ph{\chi_{ix}}=
{J_H\o2} \sum_\a  \int{dk\o2\pi} \[\Braket{\psi^\dag_{i+N+1,k\a}\psi_{i+N,k\a}}+\Braket{\psi_{i+N,k\a}^\dag \psi_{i+N+1,k\a}} \]
&
\nt
&\ph{\chi_{ix}}=
{J_H}\int{dk\o2\pi} \sum_{l=1}^{2N}n(E_l(k))\[  U_{i+N,l}(k)U^\dag_{l,i+N+1}(k)+U_{i+N,1,l}(k)U^\dag_{l,i+N+1}(k)\]
&
\\
&\chi_{iy} = {J_H\o2} \sum_\a  \int{dk\o2\pi}  2 \cos k  \Braket{f^\dag_{ik\a}f_{ik\a}}  =
{J_H} \sum_\a  \int{dk\o2\pi}\cos k  \Braket{\psi^\dag_{i+N,k\a}\psi_{i+N,k\a}} 
&
\nt
&\ph{\chi_{ix}}=
{2\,J_H}\int{dk\o2\pi} \sum_{l=1}^{2N}n(E_l(k))\cos k \; U_{i+N,l}(k)U^\dag_{l,i+N}(k) 
}

\bibliographystyle{apsrev4-1_custom}
\bibliography{tki}
\end{document}

%% file: headersDraft.tex
\usepackage[utf8]{inputenc}
\usepackage{amssymb,bm,xspace,soul} 
\usepackage{color} 
\usepackage[papersize={8.5in,11in}]{geometry}
\usepackage{natbib}
\usepackage{graphicx}
\usepackage{amsmath}
\usepackage{verbatim}
\usepackage{braket}
\usepackage{enumitem}
\usepackage{feynmp}
\usepackage{math tools} 
\usepackage{dsfont} 
\usepackage{mathrsfs} 
\usepackage{multirow}

\usepackage{setspace}

\usepackage[colorlinks=true]{hyperref} 
\hypersetup{
    bookmarks=true,         
    unicode=false,          
    pdftoolbar=true,        
    pdfmenubar=true,        
    pdffitwindow=false,     
    pdfstartview={FitH},    
    pdftitle={My title},    
    pdfauthor={Author},     
    pdfsubject={Subject},   
    pdfcreator={Creator},   
    pdfproducer={Producer}, 
    pdfkeywords={keyword1} {key2} {key3}, 
    pdfnewwindow=true,      
    colorlinks=true,       
    linkcolor=magenta, 
    citecolor=blue,        
    filecolor=magenta,      
    urlcolor=cyan           
} 

\newcommand{\ham}{\hat{H}}
\newcommand{\hamk}{\mathcal{H}}

\newcommand{\vk}{{\mathbf{k}}}
\renewcommand{\vr}{{\mathbf{r}}}

\newcommand{\V}{\mathcal{V}}

\renewcommand{\o}{\over}
\newcommand{\eq}[1]{\begin{align}#1\end{align}}
\newcommand{\eql}[1]{\begin{align}#1\end{align}}

\renewcommand{\(}{\left(}
\renewcommand{\)}{\right)}
\renewcommand{\[}{\left[}
\renewcommand{\]}{\right]}
\newcommand{\abs}[1]{\left| #1 \right|}

\newcommand{\nt}{\notag\\}

\let\v\mathbf

\newcommand{\bb}{\boldsymbol}
\newcommand{\ph}{\phantom}

\newcommand{\ep}{\epsilon}
\renewcommand{\a}{\alpha}
\renewcommand{\b}{\beta}
\renewcommand{\d}{\delta}
\newcommand{\g}{\gamma}
\newcommand{\n}{\nu}
\newcommand{\m}{\mu}

\newcommand{\s}{\sigma}

\newcommand{\lam}{\lambda}

\renewcommand{\dag}{\dagger}

\newcommand{\Zt}{\mathds{Z}_2}

\newcommand{\hvm}{\boldsymbol{\hat{\mu}}}

\newcommand{\ua}{\uparrow}
\newcommand{\da}{\downarrow}

\usepackage{tocloft} 
\addtocontents{toc}{\cftpagenumbersoff{chapter}}
\addtocontents{toc}{\cftpagenumbersoff{section}}

\setlist[enumerate]{leftmargin=*}

%% file: draft4.bbl
\begin{thebibliography}{40}%
\makeatletter
\providecommand \@ifxundefined [1]{%
 \@ifx{#1\undefined}
}%
\providecommand \@ifnum [1]{%
 \ifnum #1\expandafter \@firstoftwo
 \else \expandafter \@secondoftwo
 \fi
}%
\providecommand \@ifx [1]{%
 \ifx #1\expandafter \@firstoftwo
 \else \expandafter \@secondoftwo
 \fi
}%
\providecommand \natexlab [1]{#1}%
\providecommand \enquote  [1]{``#1''}%
\providecommand \bibnamefont  [1]{#1}%
\providecommand \bibfnamefont [1]{#1}%
\providecommand \citenamefont [1]{#1}%
\providecommand \href@noop [0]{\@secondoftwo}%
\providecommand \href [0]{\begingroup \@sanitize@url \@href}%
\providecommand \@href[1]{\@@startlink{#1}\@@href}%
\providecommand \@@href[1]{\endgroup#1\@@endlink}%
\providecommand \@sanitize@url [0]{\catcode `\\12\catcode `\$12\catcode
  `\&12\catcode `\#12\catcode `\^12\catcode `\_12\catcode `\%12\relax}%
\providecommand \@@startlink[1]{}%
\providecommand \@@endlink[0]{}%
\providecommand \url  [0]{\begingroup\@sanitize@url \@url }%
\providecommand \@url [1]{\endgroup\@href {#1}{\urlprefix }}%
\providecommand \urlprefix  [0]{URL }%
\providecommand \Eprint [0]{\href }%
\providecommand \doibase [0]{http://dx.doi.org/}%
\providecommand \selectlanguage [0]{\@gobble}%
\providecommand \bibinfo  [0]{\@secondoftwo}%
\providecommand \bibfield  [0]{\@secondoftwo}%
\providecommand \translation [1]{[#1]}%
\providecommand \BibitemOpen [0]{}%
\providecommand \bibitemStop [0]{}%
\providecommand \bibitemNoStop [0]{.\EOS\space}%
\providecommand \EOS [0]{\spacefactor3000\relax}%
\providecommand \BibitemShut  [1]{\csname bibitem#1\endcsname}%
\let\auto@bib@innerbib\@empty
\bibitem [{\citenamefont {Hasan}\ and\ \citenamefont {Kane}(2010)}]{hasan10}%
  \BibitemOpen
  \bibfield  {author} {\bibinfo {author} {\bibfnamefont {M.~Z.}\ \bibnamefont
  {Hasan}}\ and\ \bibinfo {author} {\bibfnamefont {C.~L.}\ \bibnamefont
  {Kane}},\ }\bibfield  {title} {\enquote {\bibinfo {title}
  {{\textit{Colloquium}: Topological insulators}},}\ }\href {\doibase
  10.1103/RevModPhys.82.3045} {\bibfield  {journal} {\bibinfo  {journal} {Rev.
  Mod. Phys.}\ }\textbf {\bibinfo {volume} {82}},\ \bibinfo {pages} {3045}
  (\bibinfo {year} {2010})}\BibitemShut {NoStop}%
\bibitem [{\citenamefont {Qi}\ and\ \citenamefont {Zhang}(2011)}]{qi11}%
  \BibitemOpen
  \bibfield  {author} {\bibinfo {author} {\bibfnamefont {X.-L.}\ \bibnamefont
  {Qi}}\ and\ \bibinfo {author} {\bibfnamefont {S.-C.}\ \bibnamefont {Zhang}},\
  }\bibfield  {title} {\enquote {\bibinfo {title} {{Topological insulators and
  superconductors}},}\ }\href {\doibase 10.1103/RevModPhys.83.1057} {\bibfield
  {journal} {\bibinfo  {journal} {Rev. Mod. Phys.}\ }\textbf {\bibinfo {volume}
  {83}},\ \bibinfo {pages} {1057} (\bibinfo {year} {2011})}\BibitemShut
  {NoStop}%
\bibitem [{\citenamefont {Kane}\ and\ \citenamefont {Mele}(2005)}]{qsh_kane05}%
  \BibitemOpen
  \bibfield  {author} {\bibinfo {author} {\bibfnamefont {C.~L.}\ \bibnamefont
  {Kane}}\ and\ \bibinfo {author} {\bibfnamefont {E.~J.}\ \bibnamefont
  {Mele}},\ }\bibfield  {title} {\enquote {\bibinfo {title} {{${Z}_{2}$
  Topological Order and the Quantum Spin Hall Effect}},}\ }\href {\doibase
  10.1103/PhysRevLett.95.146802} {\bibfield  {journal} {\bibinfo  {journal}
  {Phys. Rev. Lett.}\ }\textbf {\bibinfo {volume} {95}},\ \bibinfo {pages}
  {146802} (\bibinfo {year} {2005})}\BibitemShut {NoStop}%
\bibitem [{\citenamefont {Bernevig}\ \emph {et~al.}(2006)\citenamefont
  {Bernevig}, \citenamefont {Hughes},\ and\ \citenamefont {Zhang}}]{bhz}%
  \BibitemOpen
  \bibfield  {author} {\bibinfo {author} {\bibfnamefont {B.~A.}\ \bibnamefont
  {Bernevig}}, \bibinfo {author} {\bibfnamefont {T.~L.}\ \bibnamefont
  {Hughes}}, \ and\ \bibinfo {author} {\bibfnamefont {S.-C.}\ \bibnamefont
  {Zhang}},\ }\bibfield  {title} {\enquote {\bibinfo {title} {{Quantum Spin
  Hall Effect and Topological Phase Transition in HgTe Quantum Wells}},}\
  }\href {\doibase 10.1126/science.1133734} {\bibfield  {journal} {\bibinfo
  {journal} {Science}\ }\textbf {\bibinfo {volume} {314}},\ \bibinfo {pages}
  {1757} (\bibinfo {year} {2006})}\BibitemShut {NoStop}%
\bibitem [{\citenamefont {Moore}\ and\ \citenamefont
  {Balents}(2007)}]{3dTI_Moore07}%
  \BibitemOpen
  \bibfield  {author} {\bibinfo {author} {\bibfnamefont {J.~E.}\ \bibnamefont
  {Moore}}\ and\ \bibinfo {author} {\bibfnamefont {L.}~\bibnamefont
  {Balents}},\ }\bibfield  {title} {\enquote {\bibinfo {title} {Topological
  invariants of time-reversal-invariant band structures},}\ }\href {\doibase
  10.1103/PhysRevB.75.121306} {\bibfield  {journal} {\bibinfo  {journal} {Phys.
  Rev. B}\ }\textbf {\bibinfo {volume} {75}},\ \bibinfo {pages} {121306}
  (\bibinfo {year} {2007})}\BibitemShut {NoStop}%
\bibitem [{\citenamefont {Fu}\ \emph {et~al.}(2007)\citenamefont {Fu},
  \citenamefont {Kane},\ and\ \citenamefont {Mele}}]{3dTI_Fu07}%
  \BibitemOpen
  \bibfield  {author} {\bibinfo {author} {\bibfnamefont {L.}~\bibnamefont
  {Fu}}, \bibinfo {author} {\bibfnamefont {C.~L.}\ \bibnamefont {Kane}}, \ and\
  \bibinfo {author} {\bibfnamefont {E.~J.}\ \bibnamefont {Mele}},\ }\bibfield
  {title} {\enquote {\bibinfo {title} {Topological insulators in three
  dimensions},}\ }\href {\doibase 10.1103/PhysRevLett.98.106803} {\bibfield
  {journal} {\bibinfo  {journal} {Phys. Rev. Lett.}\ }\textbf {\bibinfo
  {volume} {98}},\ \bibinfo {pages} {106803} (\bibinfo {year}
  {2007})}\BibitemShut {NoStop}%
\bibitem [{\citenamefont {Roy}(2009)}]{3dTI_Roy09}%
  \BibitemOpen
  \bibfield  {author} {\bibinfo {author} {\bibfnamefont {R.}~\bibnamefont
  {Roy}},\ }\bibfield  {title} {\enquote {\bibinfo {title} {${Z}_{2}$
  classification of quantum spin hall systems: An approach using time-reversal
  invariance},}\ }\href {\doibase 10.1103/PhysRevB.79.195321} {\bibfield
  {journal} {\bibinfo  {journal} {Phys. Rev. B}\ }\textbf {\bibinfo {volume}
  {79}},\ \bibinfo {pages} {195321} (\bibinfo {year} {2009})}\BibitemShut
  {NoStop}%
\bibitem [{\citenamefont {Dzero}\ \emph {et~al.}(2010)\citenamefont {Dzero},
  \citenamefont {Sun}, \citenamefont {Galitski},\ and\ \citenamefont
  {Coleman}}]{dzero10}%
  \BibitemOpen
  \bibfield  {author} {\bibinfo {author} {\bibfnamefont {M.}~\bibnamefont
  {Dzero}}, \bibinfo {author} {\bibfnamefont {K.}~\bibnamefont {Sun}}, \bibinfo
  {author} {\bibfnamefont {V.}~\bibnamefont {Galitski}}, \ and\ \bibinfo
  {author} {\bibfnamefont {P.}~\bibnamefont {Coleman}},\ }\bibfield  {title}
  {\enquote {\bibinfo {title} {{Topological Kondo Insulators}},}\ }\href
  {\doibase 10.1103/PhysRevLett.104.106408} {\bibfield  {journal} {\bibinfo
  {journal} {Phys. Rev. Lett.}\ }\textbf {\bibinfo {volume} {104}},\ \bibinfo
  {pages} {106408} (\bibinfo {year} {2010})}\BibitemShut {NoStop}%
\bibitem [{\citenamefont {Dzero}\ \emph {et~al.}(2012)\citenamefont {Dzero},
  \citenamefont {Sun}, \citenamefont {Coleman},\ and\ \citenamefont
  {Galitski}}]{dzero12}%
  \BibitemOpen
  \bibfield  {author} {\bibinfo {author} {\bibfnamefont {M.}~\bibnamefont
  {Dzero}}, \bibinfo {author} {\bibfnamefont {K.}~\bibnamefont {Sun}}, \bibinfo
  {author} {\bibfnamefont {P.}~\bibnamefont {Coleman}}, \ and\ \bibinfo
  {author} {\bibfnamefont {V.}~\bibnamefont {Galitski}},\ }\bibfield  {title}
  {\enquote {\bibinfo {title} {{Theory of topological Kondo insulators}},}\
  }\href {\doibase 10.1103/PhysRevB.85.045130} {\bibfield  {journal} {\bibinfo
  {journal} {Phys. Rev. B}\ }\textbf {\bibinfo {volume} {85}},\ \bibinfo
  {pages} {045130} (\bibinfo {year} {2012})}\BibitemShut {NoStop}%
\bibitem [{\citenamefont {{Dzero}}\ \emph {et~al.}(2015)\citenamefont
  {{Dzero}}, \citenamefont {{Xia}}, \citenamefont {{Galitski}},\ and\
  \citenamefont {{Coleman}}}]{coleman_arcmp}%
  \BibitemOpen
  \bibfield  {author} {\bibinfo {author} {\bibfnamefont {M.}~\bibnamefont
  {{Dzero}}}, \bibinfo {author} {\bibfnamefont {J.}~\bibnamefont {{Xia}}},
  \bibinfo {author} {\bibfnamefont {V.}~\bibnamefont {{Galitski}}}, \ and\
  \bibinfo {author} {\bibfnamefont {P.}~\bibnamefont {{Coleman}}},\ }\bibfield
  {title} {\enquote {\bibinfo {title} {{Topological Kondo Insulators}},}\
  }\href@noop {} {\bibfield  {journal} {\bibinfo  {journal} {Annual Review of
  Condensed Matter Physics, to appear}\ } (\bibinfo {year} {2015})},\ \Eprint
  {http://arxiv.org/abs/1506.05635} {arXiv:1506.05635 [cond-mat.str-el]}
  \BibitemShut {NoStop}%
\bibitem [{\citenamefont {Doniach}(1977)}]{doniach77}%
  \BibitemOpen
  \bibfield  {author} {\bibinfo {author} {\bibfnamefont {S.}~\bibnamefont
  {Doniach}},\ }\bibfield  {title} {\enquote {\bibinfo {title} {{The Kondo
  lattice and weak antiferromagnetism}},}\ }\href {\doibase
  http://dx.doi.org/10.1016/0378-4363(77)90190-5} {\bibfield  {journal}
  {\bibinfo  {journal} {Physica B+C}\ }\textbf {\bibinfo {volume} {91}},\
  \bibinfo {pages} {231 } (\bibinfo {year} {1977})}\BibitemShut {NoStop}%
\bibitem [{\citenamefont {Wolgast}\ \emph {et~al.}(2013)\citenamefont
  {Wolgast}, \citenamefont {Kurdak}, \citenamefont {Sun}, \citenamefont
  {Allen}, \citenamefont {Kim},\ and\ \citenamefont {Fisk}}]{trans_Wolgast13}%
  \BibitemOpen
  \bibfield  {author} {\bibinfo {author} {\bibfnamefont {S.}~\bibnamefont
  {Wolgast}}, \bibinfo {author} {\bibfnamefont {C.}~\bibnamefont {Kurdak}},
  \bibinfo {author} {\bibfnamefont {K.}~\bibnamefont {Sun}}, \bibinfo {author}
  {\bibfnamefont {J.~W.}\ \bibnamefont {Allen}}, \bibinfo {author}
  {\bibfnamefont {D.-J.}\ \bibnamefont {Kim}}, \ and\ \bibinfo {author}
  {\bibfnamefont {Z.}~\bibnamefont {Fisk}},\ }\bibfield  {title} {\enquote
  {\bibinfo {title} {{Low-temperature surface conduction in the Kondo insulator
  SmB${}_{6}$}},}\ }\href {\doibase 10.1103/PhysRevB.88.180405} {\bibfield
  {journal} {\bibinfo  {journal} {Phys. Rev. B}\ }\textbf {\bibinfo {volume}
  {88}},\ \bibinfo {pages} {180405} (\bibinfo {year} {2013})}\BibitemShut
  {NoStop}%
\bibitem [{\citenamefont {Kim}\ \emph {et~al.}(2012)\citenamefont {Kim},
  \citenamefont {Grant},\ and\ \citenamefont {Fisk}}]{trans_Kim12}%
  \BibitemOpen
  \bibfield  {author} {\bibinfo {author} {\bibfnamefont {D.~J.}\ \bibnamefont
  {Kim}}, \bibinfo {author} {\bibfnamefont {T.}~\bibnamefont {Grant}}, \ and\
  \bibinfo {author} {\bibfnamefont {Z.}~\bibnamefont {Fisk}},\ }\bibfield
  {title} {\enquote {\bibinfo {title} {{Limit Cycle and Anomalous Capacitance
  in the Kondo Insulator ${\mathrm{SmB}}_{6}$}},}\ }\href {\doibase
  10.1103/PhysRevLett.109.096601} {\bibfield  {journal} {\bibinfo  {journal}
  {Phys. Rev. Lett.}\ }\textbf {\bibinfo {volume} {109}},\ \bibinfo {pages}
  {096601} (\bibinfo {year} {2012})}\BibitemShut {NoStop}%
\bibitem [{\citenamefont {Zhang}\ \emph {et~al.}(2013)\citenamefont {Zhang},
  \citenamefont {Butch}, \citenamefont {Syers}, \citenamefont {Ziemak},
  \citenamefont {Greene},\ and\ \citenamefont {Paglione}}]{trans_Zhang13}%
  \BibitemOpen
  \bibfield  {author} {\bibinfo {author} {\bibfnamefont {X.}~\bibnamefont
  {Zhang}}, \bibinfo {author} {\bibfnamefont {N.~P.}\ \bibnamefont {Butch}},
  \bibinfo {author} {\bibfnamefont {P.}~\bibnamefont {Syers}}, \bibinfo
  {author} {\bibfnamefont {S.}~\bibnamefont {Ziemak}}, \bibinfo {author}
  {\bibfnamefont {R.~L.}\ \bibnamefont {Greene}}, \ and\ \bibinfo {author}
  {\bibfnamefont {J.}~\bibnamefont {Paglione}},\ }\bibfield  {title} {\enquote
  {\bibinfo {title} {{Hybridization, Inter-Ion Correlation, and Surface States
  in the Kondo Insulator ${\mathrm{SmB}}_{6}$}},}\ }\href {\doibase
  10.1103/PhysRevX.3.011011} {\bibfield  {journal} {\bibinfo  {journal} {Phys.
  Rev. X}\ }\textbf {\bibinfo {volume} {3}},\ \bibinfo {pages} {011011}
  (\bibinfo {year} {2013})}\BibitemShut {NoStop}%
\bibitem [{\citenamefont {Phelan}\ \emph {et~al.}(2014)\citenamefont {Phelan},
  \citenamefont {Koohpayeh}, \citenamefont {Cottingham}, \citenamefont
  {Freeland}, \citenamefont {Leiner}, \citenamefont {Broholm},\ and\
  \citenamefont {McQueen}}]{trans_Phelan14}%
  \BibitemOpen
  \bibfield  {author} {\bibinfo {author} {\bibfnamefont {W.~A.}\ \bibnamefont
  {Phelan}}, \bibinfo {author} {\bibfnamefont {S.~M.}\ \bibnamefont
  {Koohpayeh}}, \bibinfo {author} {\bibfnamefont {P.}~\bibnamefont
  {Cottingham}}, \bibinfo {author} {\bibfnamefont {J.~W.}\ \bibnamefont
  {Freeland}}, \bibinfo {author} {\bibfnamefont {J.~C.}\ \bibnamefont
  {Leiner}}, \bibinfo {author} {\bibfnamefont {C.~L.}\ \bibnamefont {Broholm}},
  \ and\ \bibinfo {author} {\bibfnamefont {T.~M.}\ \bibnamefont {McQueen}},\
  }\bibfield  {title} {\enquote {\bibinfo {title} {{Correlation between Bulk
  Thermodynamic Measurements and the Low-Temperature-Resistance Plateau in
  ${\mathrm{SmB}}_{6}$}},}\ }\href {\doibase 10.1103/PhysRevX.4.031012}
  {\bibfield  {journal} {\bibinfo  {journal} {Phys. Rev. X}\ }\textbf {\bibinfo
  {volume} {4}},\ \bibinfo {pages} {031012} (\bibinfo {year}
  {2014})}\BibitemShut {NoStop}%
\bibitem [{\citenamefont {Kim}\ \emph {et~al.}(2013)\citenamefont {Kim},
  \citenamefont {Thomas}, \citenamefont {Grant}, \citenamefont {Botimer},
  \citenamefont {Fisk},\ and\ \citenamefont {Xia}}]{trans_Kim13}%
  \BibitemOpen
  \bibfield  {author} {\bibinfo {author} {\bibfnamefont {D.~J.}\ \bibnamefont
  {Kim}}, \bibinfo {author} {\bibfnamefont {S.}~\bibnamefont {Thomas}},
  \bibinfo {author} {\bibfnamefont {T.}~\bibnamefont {Grant}}, \bibinfo
  {author} {\bibfnamefont {J.}~\bibnamefont {Botimer}}, \bibinfo {author}
  {\bibfnamefont {Z.}~\bibnamefont {Fisk}}, \ and\ \bibinfo {author}
  {\bibfnamefont {J.}~\bibnamefont {Xia}},\ }\bibfield  {title} {\enquote
  {\bibinfo {title} {{Surface Hall effect and nonlocal transport in SmB$_6$:
  evidence for surface conduction}},}\ }\href@noop {} {\bibfield  {journal}
  {\bibinfo  {journal} {Scientific reports}\ }\textbf {\bibinfo {volume} {3}}
  (\bibinfo {year} {2013})}\BibitemShut {NoStop}%
\bibitem [{\citenamefont {Neupane}\ \emph {et~al.}(2013)\citenamefont
  {Neupane}, \citenamefont {Alidoust}, \citenamefont {Xu}, \citenamefont
  {Kondo}, \citenamefont {Ishida}, \citenamefont {Kim}, \citenamefont {Liu},
  \citenamefont {Belopolski}, \citenamefont {Jo}, \citenamefont {Chang},
  \citenamefont {Jeng}, \citenamefont {Durakiewicz}, \citenamefont {Balicas},
  \citenamefont {Lin}, \citenamefont {Bansil}, \citenamefont {Shin},
  \citenamefont {Fisk},\ and\ \citenamefont {Hasan}}]{arpes_Neupane13}%
  \BibitemOpen
  \bibfield  {author} {\bibinfo {author} {\bibfnamefont {M.}~\bibnamefont
  {Neupane}}, \bibinfo {author} {\bibfnamefont {N.}~\bibnamefont {Alidoust}},
  \bibinfo {author} {\bibfnamefont {S.-Y.}\ \bibnamefont {Xu}}, \bibinfo
  {author} {\bibfnamefont {T.}~\bibnamefont {Kondo}}, \bibinfo {author}
  {\bibfnamefont {Y.}~\bibnamefont {Ishida}}, \bibinfo {author} {\bibfnamefont
  {D.~J.}\ \bibnamefont {Kim}}, \bibinfo {author} {\bibfnamefont
  {C.}~\bibnamefont {Liu}}, \bibinfo {author} {\bibfnamefont {I.}~\bibnamefont
  {Belopolski}}, \bibinfo {author} {\bibfnamefont {Y.~J.}\ \bibnamefont {Jo}},
  \bibinfo {author} {\bibfnamefont {T.-R.}\ \bibnamefont {Chang}}, \bibinfo
  {author} {\bibfnamefont {H.-T.}\ \bibnamefont {Jeng}}, \bibinfo {author}
  {\bibfnamefont {T.}~\bibnamefont {Durakiewicz}}, \bibinfo {author}
  {\bibfnamefont {L.}~\bibnamefont {Balicas}}, \bibinfo {author} {\bibfnamefont
  {H.}~\bibnamefont {Lin}}, \bibinfo {author} {\bibfnamefont {A.}~\bibnamefont
  {Bansil}}, \bibinfo {author} {\bibfnamefont {S.}~\bibnamefont {Shin}},
  \bibinfo {author} {\bibfnamefont {Z.}~\bibnamefont {Fisk}}, \ and\ \bibinfo
  {author} {\bibfnamefont {M.~Z.}\ \bibnamefont {Hasan}},\ }\bibfield  {title}
  {\enquote {\bibinfo {title} {{Surface electronic structure of the topological
  Kondo-insulator candidate correlated electron system SmB$_6$}},}\ }\href
  {http://dx.doi.org/10.1038/ncomms3991} {\bibfield  {journal} {\bibinfo
  {journal} {Nat Commun}\ }\textbf {\bibinfo {volume} {4}} (\bibinfo {year}
  {2013})}\BibitemShut {NoStop}%
\bibitem [{\citenamefont {Jiang}\ \emph
  {et~al.}(2013{\natexlab{a}})\citenamefont {Jiang}, \citenamefont {Li},
  \citenamefont {Zhang}, \citenamefont {Sun}, \citenamefont {Chen},
  \citenamefont {Ye}, \citenamefont {Xu}, \citenamefont {Ge}, \citenamefont
  {Tan}, \citenamefont {Niu}, \citenamefont {Xia}, \citenamefont {Xie},
  \citenamefont {Li}, \citenamefont {Chen}, \citenamefont {Wen},\ and\
  \citenamefont {Feng}}]{arpes_Jiang15}%
  \BibitemOpen
  \bibfield  {author} {\bibinfo {author} {\bibfnamefont {J.}~\bibnamefont
  {Jiang}}, \bibinfo {author} {\bibfnamefont {S.}~\bibnamefont {Li}}, \bibinfo
  {author} {\bibfnamefont {T.}~\bibnamefont {Zhang}}, \bibinfo {author}
  {\bibfnamefont {Z.}~\bibnamefont {Sun}}, \bibinfo {author} {\bibfnamefont
  {F.}~\bibnamefont {Chen}}, \bibinfo {author} {\bibfnamefont {Z.~R.}\
  \bibnamefont {Ye}}, \bibinfo {author} {\bibfnamefont {M.}~\bibnamefont {Xu}},
  \bibinfo {author} {\bibfnamefont {Q.~Q.}\ \bibnamefont {Ge}}, \bibinfo
  {author} {\bibfnamefont {S.~Y.}\ \bibnamefont {Tan}}, \bibinfo {author}
  {\bibfnamefont {X.~H.}\ \bibnamefont {Niu}}, \bibinfo {author} {\bibfnamefont
  {M.}~\bibnamefont {Xia}}, \bibinfo {author} {\bibfnamefont {B.~P.}\
  \bibnamefont {Xie}}, \bibinfo {author} {\bibfnamefont {Y.~F.}\ \bibnamefont
  {Li}}, \bibinfo {author} {\bibfnamefont {X.~H.}\ \bibnamefont {Chen}},
  \bibinfo {author} {\bibfnamefont {H.~H.}\ \bibnamefont {Wen}}, \ and\
  \bibinfo {author} {\bibfnamefont {D.~L.}\ \bibnamefont {Feng}},\ }\bibfield
  {title} {\enquote {\bibinfo {title} {{Observation of possible topological
  in-gap surface states in the Kondo insulator SmB$_6$ by photoemission}},}\
  }\href {http://dx.doi.org/10.1038/ncomms4010} {\bibfield  {journal} {\bibinfo
   {journal} {Nat Commun}\ }\textbf {\bibinfo {volume} {4}} (\bibinfo {year}
  {2013}{\natexlab{a}})}\BibitemShut {NoStop}%
\bibitem [{\citenamefont {Xu}\ \emph {et~al.}(2013)\citenamefont {Xu},
  \citenamefont {Shi}, \citenamefont {Biswas}, \citenamefont {Matt},
  \citenamefont {Dhaka}, \citenamefont {Huang}, \citenamefont {Plumb},
  \citenamefont {Radovi\ifmmode~\acute{c}\else \'{c}\fi{}}, \citenamefont
  {Dil}, \citenamefont {Pomjakushina}, \citenamefont {Conder}, \citenamefont
  {Amato}, \citenamefont {Salman}, \citenamefont {Paul}, \citenamefont {Mesot},
  \citenamefont {Ding},\ and\ \citenamefont {Shi}}]{arpes_Xu13}%
  \BibitemOpen
  \bibfield  {author} {\bibinfo {author} {\bibfnamefont {N.}~\bibnamefont
  {Xu}}, \bibinfo {author} {\bibfnamefont {X.}~\bibnamefont {Shi}}, \bibinfo
  {author} {\bibfnamefont {P.~K.}\ \bibnamefont {Biswas}}, \bibinfo {author}
  {\bibfnamefont {C.~E.}\ \bibnamefont {Matt}}, \bibinfo {author}
  {\bibfnamefont {R.~S.}\ \bibnamefont {Dhaka}}, \bibinfo {author}
  {\bibfnamefont {Y.}~\bibnamefont {Huang}}, \bibinfo {author} {\bibfnamefont
  {N.~C.}\ \bibnamefont {Plumb}}, \bibinfo {author} {\bibfnamefont
  {M.}~\bibnamefont {Radovi\ifmmode~\acute{c}\else \'{c}\fi{}}}, \bibinfo
  {author} {\bibfnamefont {J.~H.}\ \bibnamefont {Dil}}, \bibinfo {author}
  {\bibfnamefont {E.}~\bibnamefont {Pomjakushina}}, \bibinfo {author}
  {\bibfnamefont {K.}~\bibnamefont {Conder}}, \bibinfo {author} {\bibfnamefont
  {A.}~\bibnamefont {Amato}}, \bibinfo {author} {\bibfnamefont
  {Z.}~\bibnamefont {Salman}}, \bibinfo {author} {\bibfnamefont {D.~M.}\
  \bibnamefont {Paul}}, \bibinfo {author} {\bibfnamefont {J.}~\bibnamefont
  {Mesot}}, \bibinfo {author} {\bibfnamefont {H.}~\bibnamefont {Ding}}, \ and\
  \bibinfo {author} {\bibfnamefont {M.}~\bibnamefont {Shi}},\ }\bibfield
  {title} {\enquote {\bibinfo {title} {{Surface and bulk electronic structure
  of the strongly correlated system SmB${}_{6}$ and implications for a
  topological Kondo insulator}},}\ }\href {\doibase 10.1103/PhysRevB.88.121102}
  {\bibfield  {journal} {\bibinfo  {journal} {Phys. Rev. B}\ }\textbf {\bibinfo
  {volume} {88}},\ \bibinfo {pages} {121102} (\bibinfo {year}
  {2013})}\BibitemShut {NoStop}%
\bibitem [{\citenamefont {Denlinger}\ \emph {et~al.}(2014)\citenamefont
  {Denlinger}, \citenamefont {Allen}, \citenamefont {Kang}, \citenamefont
  {Sun}, \citenamefont {Min}, \citenamefont {Kim},\ and\ \citenamefont
  {Fisk}}]{arpes_Denlinger14}%
  \BibitemOpen
  \bibfield  {author} {\bibinfo {author} {\bibfnamefont {J.~D.}\ \bibnamefont
  {Denlinger}}, \bibinfo {author} {\bibfnamefont {J.~W.}\ \bibnamefont
  {Allen}}, \bibinfo {author} {\bibfnamefont {J.-S.}\ \bibnamefont {Kang}},
  \bibinfo {author} {\bibfnamefont {K.}~\bibnamefont {Sun}}, \bibinfo {author}
  {\bibfnamefont {B.-I.}\ \bibnamefont {Min}}, \bibinfo {author} {\bibfnamefont
  {D.-J.}\ \bibnamefont {Kim}}, \ and\ \bibinfo {author} {\bibfnamefont
  {Z.}~\bibnamefont {Fisk}},\ }\bibfield  {title} {\enquote {\bibinfo {title}
  {{SmB$_6$ Photoemmission: Past and Present}},}\ }\href
  {http://journals.jps.jp/doi/abs/10.7566/JPSCP.3.017038} {\bibfield  {journal}
  {\bibinfo  {journal} {JPS Conf. Proc.}\ }\textbf {\bibinfo {volume} {3}},\
  \bibinfo {pages} {017038} (\bibinfo {year} {2014})}\BibitemShut {NoStop}%
\bibitem [{\citenamefont {Jiang}\ \emph
  {et~al.}(2013{\natexlab{b}})\citenamefont {Jiang}, \citenamefont {Li},
  \citenamefont {Zhang}, \citenamefont {Sun}, \citenamefont {Chen},
  \citenamefont {Ye}, \citenamefont {Xu}, \citenamefont {Ge}, \citenamefont
  {Tan}, \citenamefont {Niu}, \citenamefont {Xia}, \citenamefont {Xie},
  \citenamefont {Li}, \citenamefont {Chen}, \citenamefont {Wen},\ and\
  \citenamefont {Feng}}]{arpes_XuSpin}%
  \BibitemOpen
  \bibfield  {author} {\bibinfo {author} {\bibfnamefont {J.}~\bibnamefont
  {Jiang}}, \bibinfo {author} {\bibfnamefont {S.}~\bibnamefont {Li}}, \bibinfo
  {author} {\bibfnamefont {T.}~\bibnamefont {Zhang}}, \bibinfo {author}
  {\bibfnamefont {Z.}~\bibnamefont {Sun}}, \bibinfo {author} {\bibfnamefont
  {F.}~\bibnamefont {Chen}}, \bibinfo {author} {\bibfnamefont {Z.~R.}\
  \bibnamefont {Ye}}, \bibinfo {author} {\bibfnamefont {M.}~\bibnamefont {Xu}},
  \bibinfo {author} {\bibfnamefont {Q.~Q.}\ \bibnamefont {Ge}}, \bibinfo
  {author} {\bibfnamefont {S.~Y.}\ \bibnamefont {Tan}}, \bibinfo {author}
  {\bibfnamefont {X.~H.}\ \bibnamefont {Niu}}, \bibinfo {author} {\bibfnamefont
  {M.}~\bibnamefont {Xia}}, \bibinfo {author} {\bibfnamefont {B.~P.}\
  \bibnamefont {Xie}}, \bibinfo {author} {\bibfnamefont {Y.~F.}\ \bibnamefont
  {Li}}, \bibinfo {author} {\bibfnamefont {X.~H.}\ \bibnamefont {Chen}},
  \bibinfo {author} {\bibfnamefont {H.~H.}\ \bibnamefont {Wen}}, \ and\
  \bibinfo {author} {\bibfnamefont {D.~L.}\ \bibnamefont {Feng}},\ }\bibfield
  {title} {\enquote {\bibinfo {title} {{Observation of possible topological
  in-gap surface states in the Kondo insulator SmB$_6$ by photoemission}},}\
  }\href {http://dx.doi.org/10.1038/ncomms4010} {\bibfield  {journal} {\bibinfo
   {journal} {Nat Commun}\ }\textbf {\bibinfo {volume} {4}} (\bibinfo {year}
  {2013}{\natexlab{b}})}\BibitemShut {NoStop}%
\bibitem [{\citenamefont {Roy}\ \emph {et~al.}(2014)\citenamefont {Roy},
  \citenamefont {Sau}, \citenamefont {Dzero},\ and\ \citenamefont
  {Galitski}}]{roy14}%
  \BibitemOpen
  \bibfield  {author} {\bibinfo {author} {\bibfnamefont {B.}~\bibnamefont
  {Roy}}, \bibinfo {author} {\bibfnamefont {J.~D.}\ \bibnamefont {Sau}},
  \bibinfo {author} {\bibfnamefont {M.}~\bibnamefont {Dzero}}, \ and\ \bibinfo
  {author} {\bibfnamefont {V.}~\bibnamefont {Galitski}},\ }\bibfield  {title}
  {\enquote {\bibinfo {title} {{Surface theory of a family of topological Kondo
  insulators}},}\ }\href {\doibase 10.1103/PhysRevB.90.155314} {\bibfield
  {journal} {\bibinfo  {journal} {Phys. Rev. B}\ }\textbf {\bibinfo {volume}
  {90}},\ \bibinfo {pages} {155314} (\bibinfo {year} {2014})}\BibitemShut
  {NoStop}%
\bibitem [{\citenamefont {Nikoli\ifmmode~\acute{c}\else
  \'{c}\fi{}}(2014)}]{nikolic14}%
  \BibitemOpen
  \bibfield  {author} {\bibinfo {author} {\bibfnamefont {P.}~\bibnamefont
  {Nikoli\ifmmode~\acute{c}\else \'{c}\fi{}}},\ }\bibfield  {title} {\enquote
  {\bibinfo {title} {{Two-dimensional heavy fermions on the strongly correlated
  boundaries of Kondo topological insulators}},}\ }\href {\doibase
  10.1103/PhysRevB.90.235107} {\bibfield  {journal} {\bibinfo  {journal} {Phys.
  Rev. B}\ }\textbf {\bibinfo {volume} {90}},\ \bibinfo {pages} {235107}
  (\bibinfo {year} {2014})}\BibitemShut {NoStop}%
\bibitem [{\citenamefont {Alexandrov}\ \emph {et~al.}(2015)\citenamefont
  {Alexandrov}, \citenamefont {Coleman},\ and\ \citenamefont
  {Erten}}]{alexandrov15}%
  \BibitemOpen
  \bibfield  {author} {\bibinfo {author} {\bibfnamefont {V.}~\bibnamefont
  {Alexandrov}}, \bibinfo {author} {\bibfnamefont {P.}~\bibnamefont {Coleman}},
  \ and\ \bibinfo {author} {\bibfnamefont {O.}~\bibnamefont {Erten}},\
  }\bibfield  {title} {\enquote {\bibinfo {title} {{Kondo Breakdown in
  Topological Kondo Insulators}},}\ }\href {\doibase
  10.1103/PhysRevLett.114.177202} {\bibfield  {journal} {\bibinfo  {journal}
  {Phys. Rev. Lett.}\ }\textbf {\bibinfo {volume} {114}},\ \bibinfo {pages}
  {177202} (\bibinfo {year} {2015})}\BibitemShut {NoStop}%
\bibitem [{\citenamefont {Iaconis}\ and\ \citenamefont
  {Balents}(2015)}]{iaconis15}%
  \BibitemOpen
  \bibfield  {author} {\bibinfo {author} {\bibfnamefont {J.}~\bibnamefont
  {Iaconis}}\ and\ \bibinfo {author} {\bibfnamefont {L.}~\bibnamefont
  {Balents}},\ }\bibfield  {title} {\enquote {\bibinfo {title} {{Many-body
  effects in topological Kondo insulators}},}\ }\href {\doibase
  10.1103/PhysRevB.91.245127} {\bibfield  {journal} {\bibinfo  {journal} {Phys.
  Rev. B}\ }\textbf {\bibinfo {volume} {91}},\ \bibinfo {pages} {245127}
  (\bibinfo {year} {2015})}\BibitemShut {NoStop}%
\bibitem [{\citenamefont {Alexandrov}\ \emph {et~al.}(2013)\citenamefont
  {Alexandrov}, \citenamefont {Dzero},\ and\ \citenamefont
  {Coleman}}]{alexandrov13}%
  \BibitemOpen
  \bibfield  {author} {\bibinfo {author} {\bibfnamefont {V.}~\bibnamefont
  {Alexandrov}}, \bibinfo {author} {\bibfnamefont {M.}~\bibnamefont {Dzero}}, \
  and\ \bibinfo {author} {\bibfnamefont {P.}~\bibnamefont {Coleman}},\
  }\bibfield  {title} {\enquote {\bibinfo {title} {{Cubic Topological Kondo
  Insulators}},}\ }\href {\doibase 10.1103/PhysRevLett.111.226403} {\bibfield
  {journal} {\bibinfo  {journal} {Phys. Rev. Lett.}\ }\textbf {\bibinfo
  {volume} {111}},\ \bibinfo {pages} {226403} (\bibinfo {year}
  {2013})}\BibitemShut {NoStop}%
\bibitem [{\citenamefont {Lu}\ \emph {et~al.}(2013)\citenamefont {Lu},
  \citenamefont {Zhao}, \citenamefont {Weng}, \citenamefont {Fang},\ and\
  \citenamefont {Dai}}]{lu13}%
  \BibitemOpen
  \bibfield  {author} {\bibinfo {author} {\bibfnamefont {F.}~\bibnamefont
  {Lu}}, \bibinfo {author} {\bibfnamefont {J.}~\bibnamefont {Zhao}}, \bibinfo
  {author} {\bibfnamefont {H.}~\bibnamefont {Weng}}, \bibinfo {author}
  {\bibfnamefont {Z.}~\bibnamefont {Fang}}, \ and\ \bibinfo {author}
  {\bibfnamefont {X.}~\bibnamefont {Dai}},\ }\bibfield  {title} {\enquote
  {\bibinfo {title} {{Correlated Topological Insulators with Mixed Valence}},}\
  }\href {\doibase 10.1103/PhysRevLett.110.096401} {\bibfield  {journal}
  {\bibinfo  {journal} {Phys. Rev. Lett.}\ }\textbf {\bibinfo {volume} {110}},\
  \bibinfo {pages} {096401} (\bibinfo {year} {2013})}\BibitemShut {NoStop}%
\bibitem [{\citenamefont {Balents}(2010)}]{balents_review}%
  \BibitemOpen
  \bibfield  {author} {\bibinfo {author} {\bibfnamefont {L.}~\bibnamefont
  {Balents}},\ }\bibfield  {title} {\enquote {\bibinfo {title} {{Spin liquids
  in frustrated magnets}},}\ }\href {http://dx.doi.org/10.1038/nature08917}
  {\bibfield  {journal} {\bibinfo  {journal} {Nature}\ }\textbf {\bibinfo
  {volume} {464}},\ \bibinfo {pages} {199} (\bibinfo {year}
  {2010})}\BibitemShut {NoStop}%
\bibitem [{\citenamefont {Senthil}\ \emph {et~al.}(2003)\citenamefont
  {Senthil}, \citenamefont {Sachdev},\ and\ \citenamefont {Vojta}}]{senthil03}%
  \BibitemOpen
  \bibfield  {author} {\bibinfo {author} {\bibfnamefont {T.}~\bibnamefont
  {Senthil}}, \bibinfo {author} {\bibfnamefont {S.}~\bibnamefont {Sachdev}}, \
  and\ \bibinfo {author} {\bibfnamefont {M.}~\bibnamefont {Vojta}},\ }\bibfield
   {title} {\enquote {\bibinfo {title} {{Fractionalized Fermi Liquids}},}\
  }\href {\doibase 10.1103/PhysRevLett.90.216403} {\bibfield  {journal}
  {\bibinfo  {journal} {Phys. Rev. Lett.}\ }\textbf {\bibinfo {volume} {90}},\
  \bibinfo {pages} {216403} (\bibinfo {year} {2003})}\BibitemShut {NoStop}%
\bibitem [{\citenamefont {Senthil}\ \emph {et~al.}(2004)\citenamefont
  {Senthil}, \citenamefont {Vojta},\ and\ \citenamefont {Sachdev}}]{senthil04}%
  \BibitemOpen
  \bibfield  {author} {\bibinfo {author} {\bibfnamefont {T.}~\bibnamefont
  {Senthil}}, \bibinfo {author} {\bibfnamefont {M.}~\bibnamefont {Vojta}}, \
  and\ \bibinfo {author} {\bibfnamefont {S.}~\bibnamefont {Sachdev}},\
  }\bibfield  {title} {\enquote {\bibinfo {title} {{Weak magnetism and
  non-Fermi liquids near heavy-fermion critical points}},}\ }\href {\doibase
  10.1103/PhysRevB.69.035111} {\bibfield  {journal} {\bibinfo  {journal} {Phys.
  Rev. B}\ }\textbf {\bibinfo {volume} {69}},\ \bibinfo {pages} {035111}
  (\bibinfo {year} {2004})}\BibitemShut {NoStop}%
\bibitem [{\citenamefont {Anderson}(1961)}]{anderson61}%
  \BibitemOpen
  \bibfield  {author} {\bibinfo {author} {\bibfnamefont {P.~W.}\ \bibnamefont
  {Anderson}},\ }\bibfield  {title} {\enquote {\bibinfo {title} {{Localized
  Magnetic States in Metals}},}\ }\href {\doibase 10.1103/PhysRev.124.41}
  {\bibfield  {journal} {\bibinfo  {journal} {Phys. Rev.}\ }\textbf {\bibinfo
  {volume} {124}},\ \bibinfo {pages} {41} (\bibinfo {year} {1961})}\BibitemShut
  {NoStop}%
\bibitem [{\citenamefont {Schrieffer}\ and\ \citenamefont
  {Wolff}(1966)}]{schriefferwolff}%
  \BibitemOpen
  \bibfield  {author} {\bibinfo {author} {\bibfnamefont {J.~R.}\ \bibnamefont
  {Schrieffer}}\ and\ \bibinfo {author} {\bibfnamefont {P.~A.}\ \bibnamefont
  {Wolff}},\ }\bibfield  {title} {\enquote {\bibinfo {title} {{Relation between
  the Anderson and Kondo Hamiltonians}},}\ }\href {\doibase
  10.1103/PhysRev.149.491} {\bibfield  {journal} {\bibinfo  {journal} {Phys.
  Rev.}\ }\textbf {\bibinfo {volume} {149}},\ \bibinfo {pages} {491} (\bibinfo
  {year} {1966})}\BibitemShut {NoStop}%
\bibitem [{\citenamefont {Coqblin}\ and\ \citenamefont
  {Schrieffer}(1969)}]{coqblinschrieffer}%
  \BibitemOpen
  \bibfield  {author} {\bibinfo {author} {\bibfnamefont {B.}~\bibnamefont
  {Coqblin}}\ and\ \bibinfo {author} {\bibfnamefont {J.~R.}\ \bibnamefont
  {Schrieffer}},\ }\bibfield  {title} {\enquote {\bibinfo {title} {Exchange
  interaction in alloys with cerium impurities},}\ }\href {\doibase
  10.1103/PhysRev.185.847} {\bibfield  {journal} {\bibinfo  {journal} {Phys.
  Rev.}\ }\textbf {\bibinfo {volume} {185}},\ \bibinfo {pages} {847} (\bibinfo
  {year} {1969})}\BibitemShut {NoStop}%
\bibitem [{\citenamefont {Bernevig}\ and\ \citenamefont
  {Hughes}(2013)}]{bernevig}%
  \BibitemOpen
  \bibfield  {author} {\bibinfo {author} {\bibfnamefont {B.~A.}\ \bibnamefont
  {Bernevig}}\ and\ \bibinfo {author} {\bibfnamefont {T.~L.}\ \bibnamefont
  {Hughes}},\ }\href {https://books.google.com/books?id=wOn7JHSSxrsC} {\emph
  {\bibinfo {title} {{Topological Insulators and Topological
  Superconductors}}}}\ (\bibinfo  {publisher} {Princeton University Press},\
  \bibinfo {year} {2013})\BibitemShut {NoStop}%
\bibitem [{\citenamefont {Fu}\ and\ \citenamefont {Kane}(2007)}]{fukane}%
  \BibitemOpen
  \bibfield  {author} {\bibinfo {author} {\bibfnamefont {L.}~\bibnamefont
  {Fu}}\ and\ \bibinfo {author} {\bibfnamefont {C.~L.}\ \bibnamefont {Kane}},\
  }\bibfield  {title} {\enquote {\bibinfo {title} {{Topological insulators with
  inversion symmetry}},}\ }\href {\doibase 10.1103/PhysRevB.76.045302}
  {\bibfield  {journal} {\bibinfo  {journal} {Phys. Rev. B}\ }\textbf {\bibinfo
  {volume} {76}},\ \bibinfo {pages} {045302} (\bibinfo {year}
  {2007})}\BibitemShut {NoStop}%
\bibitem [{\citenamefont {Paul}\ \emph {et~al.}(2008)\citenamefont {Paul},
  \citenamefont {P\'epin},\ and\ \citenamefont {Norman}}]{paul08}%
  \BibitemOpen
  \bibfield  {author} {\bibinfo {author} {\bibfnamefont {I.}~\bibnamefont
  {Paul}}, \bibinfo {author} {\bibfnamefont {C.}~\bibnamefont {P\'epin}}, \
  and\ \bibinfo {author} {\bibfnamefont {M.~R.}\ \bibnamefont {Norman}},\
  }\bibfield  {title} {\enquote {\bibinfo {title} {{Multiscale fluctuations
  near a Kondo breakdown quantum critical point}},}\ }\href {\doibase
  10.1103/PhysRevB.78.035109} {\bibfield  {journal} {\bibinfo  {journal} {Phys.
  Rev. B}\ }\textbf {\bibinfo {volume} {78}},\ \bibinfo {pages} {035109}
  (\bibinfo {year} {2008})}\BibitemShut {NoStop}%
\bibitem [{\citenamefont {Haldane}(1988)}]{haldane}%
  \BibitemOpen
  \bibfield  {author} {\bibinfo {author} {\bibfnamefont {F.~D.~M.}\
  \bibnamefont {Haldane}},\ }\bibfield  {title} {\enquote {\bibinfo {title}
  {O(3) nonlinear $\ensuremath{\sigma}$ model and the topological distinction
  between integer- and half-integer-spin antiferromagnets in two dimensions},}\
  }\href {\doibase 10.1103/PhysRevLett.61.1029} {\bibfield  {journal} {\bibinfo
   {journal} {Phys. Rev. Lett.}\ }\textbf {\bibinfo {volume} {61}},\ \bibinfo
  {pages} {1029} (\bibinfo {year} {1988})}\BibitemShut {NoStop}%
\bibitem [{\citenamefont {Giamarchi}(2003)}]{giamarchi03}%
  \BibitemOpen
  \bibfield  {author} {\bibinfo {author} {\bibfnamefont {T.}~\bibnamefont
  {Giamarchi}},\ }\href {https://books.google.ca/books?id=GVeuKZLGMZ0C} {\emph
  {\bibinfo {title} {Quantum Physics in One Dimension}}},\ International Series
  of Monographs on Physics\ (\bibinfo  {publisher} {Clarendon Press},\ \bibinfo
  {year} {2003})\BibitemShut {NoStop}%
\bibitem [{\citenamefont {Tan}\ \emph {et~al.}(2015)\citenamefont {Tan},
  \citenamefont {Hsu}, \citenamefont {Zeng}, \citenamefont {Hatnean},
  \citenamefont {Harrison}, \citenamefont {Zhu}, \citenamefont {Hartstein},
  \citenamefont {Kiourlappou}, \citenamefont {Srivastava}, \citenamefont
  {Johannes}, \citenamefont {Murphy}, \citenamefont {Park}, \citenamefont
  {Balicas}, \citenamefont {Lonzarich}, \citenamefont {Balakrishnan},\ and\
  \citenamefont {Sebastian}}]{QO_suchitra15}%
  \BibitemOpen
  \bibfield  {author} {\bibinfo {author} {\bibfnamefont {B.~S.}\ \bibnamefont
  {Tan}}, \bibinfo {author} {\bibfnamefont {Y.-T.}\ \bibnamefont {Hsu}},
  \bibinfo {author} {\bibfnamefont {B.}~\bibnamefont {Zeng}}, \bibinfo {author}
  {\bibfnamefont {M.~C.}\ \bibnamefont {Hatnean}}, \bibinfo {author}
  {\bibfnamefont {N.}~\bibnamefont {Harrison}}, \bibinfo {author}
  {\bibfnamefont {Z.}~\bibnamefont {Zhu}}, \bibinfo {author} {\bibfnamefont
  {M.}~\bibnamefont {Hartstein}}, \bibinfo {author} {\bibfnamefont
  {M.}~\bibnamefont {Kiourlappou}}, \bibinfo {author} {\bibfnamefont
  {A.}~\bibnamefont {Srivastava}}, \bibinfo {author} {\bibfnamefont {M.~D.}\
  \bibnamefont {Johannes}}, \bibinfo {author} {\bibfnamefont {T.~P.}\
  \bibnamefont {Murphy}}, \bibinfo {author} {\bibfnamefont {J.-H.}\
  \bibnamefont {Park}}, \bibinfo {author} {\bibfnamefont {L.}~\bibnamefont
  {Balicas}}, \bibinfo {author} {\bibfnamefont {G.~G.}\ \bibnamefont
  {Lonzarich}}, \bibinfo {author} {\bibfnamefont {G.}~\bibnamefont
  {Balakrishnan}}, \ and\ \bibinfo {author} {\bibfnamefont {S.~E.}\
  \bibnamefont {Sebastian}},\ }\bibfield  {title} {\enquote {\bibinfo {title}
  {{Unconventional Fermi surface in an insulating state}},}\ }\href {\doibase
  10.1126/science.aaa7974} {\bibfield  {journal} {\bibinfo  {journal}
  {Science}\ }\textbf {\bibinfo {volume} {349}},\ \bibinfo {pages} {287}
  (\bibinfo {year} {2015})}\BibitemShut {NoStop}%
\bibitem [{\citenamefont {{Knolle}}\ and\ \citenamefont
  {{Cooper}}(2015)}]{Cooper15}%
  \BibitemOpen
  \bibfield  {author} {\bibinfo {author} {\bibfnamefont {J.}~\bibnamefont
  {{Knolle}}}\ and\ \bibinfo {author} {\bibfnamefont {N.~R.}\ \bibnamefont
  {{Cooper}}},\ }\bibfield  {title} {\enquote {\bibinfo {title} {{Quantum
  oscillations without a Fermi surface and the anomalous de Haas-van Alphen
  effect}},}\ }\href@noop {} {\bibfield  {journal} {\bibinfo  {journal} {ArXiv
  e-prints}\ } (\bibinfo {year} {2015})},\ \Eprint
  {http://arxiv.org/abs/1507.00885} {arXiv:1507.00885 [cond-mat.str-el]}
  \BibitemShut {NoStop}%
\end{thebibliography}%
